\documentclass[twocolumn,aps,prd,superscriptaddress,preprintnumbers,nofootinbib,10pt]{revtex4-2}
\usepackage{amsfonts}
\usepackage[tbtags]{amsmath}
\usepackage{amssymb}
\usepackage{color}
\usepackage{courier}
\usepackage{dsfont}
\usepackage{euscript}
\usepackage{epsfig}
\usepackage{epstopdf}
\usepackage{float}
\usepackage{graphics}
\usepackage{graphicx}
\usepackage{hyperref}
\usepackage[utf8]{inputenc}
\usepackage{latexsym}
\usepackage{MnSymbol}
\usepackage{pifont}
\usepackage{physics}
\usepackage{slashed}
\usepackage{subfigure}
\usepackage{tikz-feynman}
\usepackage[normalem]{ulem}
\usepackage{url}
\usepackage{verbatim}
\usepackage{widetable}

\allowdisplaybreaks

\hypersetup{
colorlinks=true,
citecolor=blue,
citebordercolor=red,
linktoc=all,
linkcolor=blue,
urlcolor=blue
}

\def \ie{{\it i.e.}}
\def \pt{\partial}

\def \and{\textmd{and}}
\def \nn{\nonumber} 
 
\def \T{\textmd{T}}
\def \GN{G_{\textmd{N}}}

\def \gf{\textmd{g.f.}}

\def \SU{\textmd{SU}}
\def \U{\textmd{U}}

\def \Y{\textmd{Y}}
\def \L{\textmd{L}}
\def \R{\textmd{R}}
\def \SM{\textmd{SM}}

\def \D{\textmd{D}}
\def \H{\textmd{H}}
\def \A{\textmd{A}}
\def \B{\textmd{B}}
\def \W{\textmd{W}}
\def \S{\textmd{S}}

\newcommand{\al}[1]{\begin{align}#1\end{align}}

\def \be{\begin{equation}}
\def \ee{\end{equation}}

\def \bea{\begin{eqnarray}}
\def \eea{\end{eqnarray}}

\graphicspath{{./Figures/}}
\newbox{\ORCIDicon}
\sbox{\ORCIDicon}{\large\includegraphics[width=0.8em]{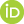}}

\begin{document}

\title{Ruling out models of vector dark matter in asymptotically safe quantum gravity
}

\author{Gustavo P. de Brito\,\href{https://orcid.org/0000-0003-2240-528X}{\usebox{\ORCIDicon}}}
\email{gustavo@cp3.sdu.dk}
\affiliation{CP3-Origins, University of Southern Denmark, Campusvej 55, DK-5230 Odense M, Denmark}
\author{Astrid Eichhorn\, \href{https://orcid.org/0000-0003-4458-1495}{\usebox{\ORCIDicon}}}
\email{eichhorn@cp3.sdu.dk}
\affiliation{CP3-Origins, University of Southern Denmark, Campusvej 55, DK-5230 Odense M, Denmark}
\author{Mads T. Frandsen\, \href{https://orcid.org/0000-0003-2061-562X}{\usebox{\ORCIDicon}}} 
\email{frandsen@cp3.sdu.dk}
\affiliation{CP3-Origins, University of Southern Denmark, Campusvej 55, DK-5230 Odense M, Denmark}
\author{Martin Rosenlyst\, \href{https://orcid.org/0000-0003-1245-7748}{\usebox{\ORCIDicon}}}
\email{martin.jorgensen@physics.ox.ac.uk}
\affiliation{Rudolf Peierls Centre for Theoretical Physics, University of Oxford, 1 Keble Road, Oxford OX1	3NP, United Kingdom}
\author{Mattias E. Thing\,\href{https://orcid.org/0000-0003-3523-9700}{\usebox{\ORCIDicon}}} 
\email{thing@cp3.sdu.dk}
\affiliation{CP3-Origins, University of Southern Denmark, Campusvej 55, DK-5230 Odense M, Denmark}
\author{Arthur F. Vieira\,\href{https://orcid.org/0000-0003-2897-2437}{\usebox{\ORCIDicon}}} 
\email{afv@cp3.sdu.dk, arthurfv@id.uff.br}
\affiliation{CP3-Origins, University of Southern Denmark, Campusvej 55, DK-5230 Odense M, Denmark}
\affiliation{Instituto de F\'isica, Universidade Federal Fluminense, Campus da Praia Vermelha, Av. Litor\^anea s/n, 24210-346, Niter\'oi, Rio de Janeiro, Brazil}

\begin{abstract}
The nature of dark matter is a problem with too many potential solutions. We investigate whether a consistent embedding into quantum gravity can decimate the number of solutions to the dark-matter problem. Concretely, we focus on a hidden sector composed of a gauge field and a charged scalar, with gauge group U(1)$_{\textmd{D}}$ or SU(2)$_\textmd{D}$. The gauge field is the dark matter candidate, if the gauge symmetry is broken spontaneously. Phenomenological constraints on the couplings in this model arise from requiring that the correct dark matter relic density is produced via thermal freeze-out and that recent bounds from direct-detection experiments are respected. We find that the consistent embedding into asymptotically safe quantum gravity gives rise to additional constraints on the couplings at the Planck scale, from which we calculate corresponding constraints at low energy scales. We discover that phenomenological constraints cannot be satisfied simultaneously with theoretical constraints from asymptotically safe quantum gravity, ruling out these dark matter models. 
\end{abstract}

\maketitle

\section{Introduction \label{Sec:Intro}}
\subsection{Motivation for connecting quantum gravity\\and dark matter}

Fundamental physics faces several profound challenges. One is to understand the quantum nature of gravity and another is to understand the true nature of the dark matter (DM). At a first glance, these challenges appear unrelated, because they are associated to very different energy scales. Quantum gravity is typically assumed to become dynamically important at energies of $E \approx M_{\rm Planck} = 10^{19}\, \rm GeV$. DM candidates span a huge range in masses \cite{Cooley:2022ufh,Arbey:2021gdg,Boveia:2022syt,Akerib:2022ort,Essig:2022dfa, Carney:2022gse,Baxter:2022dkm, Aramaki:2022zpw,Ando:2022kzd, Kahn:2022kae,Leane:2022bfm,Mitridate:2022tnv}, but most proposals focus on energy scales far below the Planck scale, with the typical mass scales for weakly interacting massive particles (WIMPs) in the GeV-TeV range \cite{Steigman:1985cco}. In this paper, we advocate that much can be learned about both quantum gravity and DM, if we consider both challenges simultaneously. We support this claim by providing a concrete example.\\
The key idea underlying our paper is that the interplay of quantum gravity with  dark (and visible) matter\footnote{In this context, by matter we refer to all fields except the metric; \ie, gauge fields are part of the matter sector.}  imprints structures on and constrains the couplings of the matter sector at the Planck scale, see \cite{Eichhorn:2022gku} for a review. The renormalization group (RG) then acts as a lever arm that translates these structures at tiny length scales (high energies) to structures at large length scales (low energies). Thereby, two important goals are achieved: first, by generating predictions for the interaction structure of dark and visible matter, the quantum gravity theory becomes testable by current observations. Second, by predicting the interaction structure of the DM, the huge space of DM models is reduced and, as we will show, phenomenologically viable DM models are ruled out on theoretical grounds. Both goals - making quantum gravity testable and making DM models more predictive - are critical in order to make progress in our understanding of our Universe.\\
This paper supports the idea that such progress can be made, if we overcome the division between quantum-gravity research and DM research, and consider quantum gravity and DM together in a multiscale setup. In such a multiscale setup, theoretical constraints at the Planck scale are combined with phenomenological and observational constraints at lower scales and the resulting theory of DM and quantum gravity is significantly more predictive than a theory of DM on its own.\\ 

In the present paper, we explore this general idea in the context of asymptotically safe quantum gravity and weakly interacting massive particle (WIMP) dark matter.

We have compelling evidence from cosmological and astrophysical observations ranging from the cosmic microwave background to dwarf galaxies~\cite{Planck:2018vyg,Zwicky:1933ren,Rubin:1983tro} that the majority of the matter density in the Universe is nonbaryonic DM with no particle candidates in the Standard Model (SM) able to account for this DM content. The most studied DM particle candidate has been the WIMP. However, the WIMP paradigm has become highly challenged by the absence of expected WIMP signals in DM particle search experiments; both direct, indirect, and collider experiment~\cite{ Billard:2021uyg,Cooley:2022ufh,ATLAS:2022hsp,Kalinowski:2022fot,CMS:2022sfl,CMS:2018fux}. Therefore, there is now high motivation to study the remaining parts of the WIMP parameter space as well as alternatives to the WIMP paradigm and in particular new ways to constrain the vast parameter space of such alternative paradigms. 

Asymptotically safe gravity is a quantum field theory of the metric. The problem of perturbative nonrenormalizability is solved by an ultraviolet (UV) fixed point in the renormalization group (RG) evolution of the theory. Because of the fixed point, quantum scale symmetry is realized at trans-Planckian scales. At the Planck scale, the theory departs from the scale-symmetric regime along one of the relevant perturbations of the fixed point, and gravitational couplings, such as the Newton coupling and cosmological constant, evolve to their measured low-energy values \cite{Reuter:2001ag,Gubitosi:2018gsl}. This scenario for quantum gravity is supported by numerous studies \cite{Reuter:1996cp,Lauscher:2001ya,Reuter:2001ag,Litim:2003vp,Codello:2006in,Codello:2007bd,Machado:2007ea,Codello:2008vh,Benedetti:2009rx,Benedetti:2010nr,Manrique:2011jc,Falls:2013bv,Codello:2013fpa,Christiansen:2014raa,Becker:2014qya,Christiansen:2015rva,Ohta:2015efa,Gies:2016con,Biemans:2016rvp,Denz:2016qks,Gonzalez-Martin:2017gza,Knorr:2017mhu,Christiansen:2017bsy,Bosma:2019aiu,Knorr:2019atm,Falls:2020qhj,Kluth:2020bdv,Knorr:2020ckv,Bonanno:2021squ,Fehre:2021eob,Knorr:2021slg,Knorr:2021niv,Baldazzi:2021orb,Mitchell:2021qjr,Sen:2021ffc,Kluth:2022vnq,Knorr:2023usb,Pawlowski:2023dda,Saueressig:2023tfy}, 
see \cite{Eichhorn:2018yfc,Eichhorn:2022gku,Pawlowski:2023gym,Saueressig:2023irs, Knorr:2022dsx,Pawlowski:2020qer} for recent reviews, \cite{percacci2017introduction,Reuter:2019byg} for books, \cite{Nagy:2012ef,Reichert:2020mja,Eichhorn:2020mte} for lecture notes, and \cite{Donoghue:2019clr,Bonanno:2020bil} for a critical discussion of the state of the field.

The asymptotic-safety paradigm has been extended to include the SM matter fields, see \cite{Eichhorn:2022jqj,Eichhorn:2022gku,Eichhorn:2023xee} for reviews.
 There is robust evidence that gravity remains asymptotically safe with the inclusion of fluctuations of Standard Model fields \cite{Dona:2013qba,Meibohm:2015twa,Biemans:2017zca,Christiansen:2017cxa,Alkofer:2018fxj,Eichhorn:2018ydy,Wetterich:2019zdo,deBrito:2020xhy}. The interplay of gravity with matter provides a mechanism that fixes couplings in the Standard Model from first principles \cite{Shaposhnikov:2009pv,Harst:2011zx,Eichhorn:2017eht,Christiansen:2017gtg,Eichhorn:2017ylw,Eichhorn:2017lry, Eichhorn:2018whv,DeBrito:2019gdd,Alkofer:2020vtb,deBrito:2022vbr,Pastor-Gutierrez:2022nki}. The interplay of asymptotic safety with scalars \cite{Narain:2009fy,Oda:2015sma,Percacci:2015wwa, Labus:2015ska, Dona:2015tnf,Eichhorn:2018akn,Ali:2020znq,deBrito:2021pyi,Laporte:2021kyp,Eichhorn:2022ngh,deBrito:2023myf}, fermions \cite{Eichhorn:2011pc,Dona:2012am, Meibohm:2016mkp, Eichhorn:2016vvy, Eichhorn:2018nda,DeBrito:2019rrh, Daas:2020dyo, Daas:2021abx} and gauge fields \cite{Folkerts:2011jz,Harst:2011zx,Christiansen:2017gtg,Christiansen:2017cxa,Eichhorn:2019yzm,Eichhorn:2021qet} has also been investigated separately.
Extensions beyond the Standard Model have also been explored~\cite{Eichhorn:2017als,Eichhorn:2019ybe,Reichert:2019car,Eichhorn:2020sbo,Hamada:2020vnf,Eichhorn:2020kca,Kowalska:2020gie,Kowalska:2020zve,Eichhorn:2021tsx,deBrito:2021akp,Kowalska:2022ypk,Eichhorn:2022vgp,Boos:2022jvc,Boos:2022pyq,Chikkaballi:2022urc,Chikkaballi:2023cce,Kotlarski:2023mmr,Eichhorn:2023gat}.
Dark matter was studied for the first time in \cite{Eichhorn:2017als}, where it was discovered that the Higgs portal to a single, uncharged dark scalar must vanish, ruling out this simplest WIMP candidate. Extended WIMP models have been considered in 
\cite{Reichert:2019car,Hamada:2020vnf,Eichhorn:2020kca,Eichhorn:2020sbo,Kowalska:2020zve,Boos:2022pyq}
 and axionlike dark matter has been considered as well \cite{deBrito:2021akp}. In the present paper, we start from a WIMP model that is phenomenologically viable, as explored in depth in \cite{Frandsen:2022klh}, and investigate whether or not it is compatible with asymptotic safety.

This paper is structured as follows:  In Sec.~\ref{Sec:Models} we define the dark-matter-gravity models that we explore by introducing field content, interactions and symmetries. In Sec.~\ref{sec:betas} we provide an overview of the methodology to calculate beta functions and list the beta functions of our model.
In Sec.~\ref{sec:results} we provide the results of our analysis and present conclusions and an outlook in Sec.~\ref{sec:conclusions}. In the Appendix, we provide additional details regarding the choice of gauge of our calculation in the gravitational and the matter sector.

%

\section{SM-Gravity-Dark Matter systems: definition of the models \label{Sec:Models}}
We consider two extensions of the SM coupled to gravity: $\U(1)_\D$ and $\SU(2)_\D$ hidden DM models with vector DM candidates and a new SM singlet scalar $S$ that is charged under the dark gauge group. Schematically, the classical gauge-fixed action for the gravity-matter dynamics reads
\be
S=S_{\textmd{grav}}+S_{\SM}^0+S_{\textmd{DM}}.
\ee
For the SM subsystem, we consider the gauge interactions of the SM gauge group $\U(1)_\Y\times\SU(2)_{\L}\times\SU(3)_{\textmd{C}}$ coupled to the quarks and leptons. In particular,
for the Yukawa sector, we consider an approximation where only the top and bottom quarks have nonvanishing (real) Yukawa couplings. We denote by $S_{\SM}^0$ the SM action without the Higgs potential, and $g_\Y=\sqrt{\frac{3}{5}}g_1$, $g_2$ and $g_3$ are the respective SM gauge couplings, which are dimensionless in four dimensions. Explicit forms and conventions for our system are displayed in Appendix \ref{App::gauge-fixed_action}.

We first extend the Standard Model by a dark complex scalar $S$ charged under a $\U(1)_\D$ gauge symmetry with gauge boson $V_\mu$.
Explicitly, the gauge-fixed action reads~\cite{YaserAyazi:2019caf}\footnote{ Throughout the work we use the shorthand notations for volume integrals of the full and background metrics, respectively, $\int_x= \int \dd^4x\, \sqrt{g}$ and $\int_{\bar{x}}=\int \dd^4x\, \sqrt{\bar{g}}$.}
\begin{align}\label{eq::actionU1}
	S_{\textmd{DM}}^{\U(1)_\D}&=\int_x\,\bigg[\frac{1}{4}V_{\mu\nu}V^{\mu\nu}+(D_\mu S)^*(D^\mu S)+V(\Phi,S)\bigg]\nn\\[1ex]
	&+\frac{1}{2\xi_{\D}}\int_{\bar{x}}\left(\pt_\alpha V^{\alpha} \right)^2,
\end{align}
 where $V_{\mu\nu}=\partial_\mu V_\nu-\partial_\nu V_\mu$ is the field strength tensor for the U(1)$_\D$ gauge field and
 \al{
 	D_\mu S=\partial_\mu S-i g_{\D} V_\mu S,
 }
 in which $g_{\D}$ is the U(1)$_\D$ gauge coupling and $\Phi$ is the $\SU(2)_\L$ Higgs doublet. For explicit computations, we adopt the Landau-gauge limit, \ie, $\xi_{\D}\rightarrow 0$ and the corresponding Faddeev-Popov ghosts only contribute to the flow of the gravitational couplings. An additional unbroken $\mathbb{Z}_{2}$ symmetry is present, under which the dark vector boson transforms as
 	\be
 V_\mu \mapsto -V_\mu,
 \ee
 while all other fields are even. This symmetry ensures the stability of the dark sector and prohibits a kinetic mixing between the dark vector $V_\mu$ and the gauge field of the hypercharge sector of the SM, \ie, a term of the form $V_{\mu\nu}B^{\mu\nu}$, where $B_{\mu\nu}=\partial_\mu B_\nu-\partial_\nu B_\mu$.
 
 The scalar potential supplemented by a portal interaction with the Higgs doublet is given by \cite{Frandsen:2022klh}
 \al{\label{eq::U1potential}
 	V(\Phi,S)&=m^2_{\H}\Phi_i^\dagger \Phi_i+\frac{\lambda_\H}{6}(\Phi_i^\dagger \Phi_i)^2+m^2_{\S}S^*S\nn\\[1ex]
 	&+\frac{\lambda_\S}{6}(S^*S)^2+2\lambda_{p}(\Phi_i^\dagger \Phi_i)(S^*S),
 }
 where the quartic couplings $\lambda_\H$, $\lambda_\S$ and $\lambda_p$ are dimensionless in four dimensions.

 The second model to be considered is the non-Abelian $\SU(2)_\D$ extension. In this model, the DM candidate comprises a $\SU(2)_\D$ vector triplet $V^a_{\mu}$ with $a=1,2,3$, alongside a complex $\SU(2)_\D$ doublet $S_i$ with $i=1,2$. Explicitly, the gauge-fixed action is given, in this case, by~\cite{Hambye:2013dgv}
\begin{align}\label{eq::actionSU2}
	S_{\textmd{DM}}^{\SU(2)_\D}&=
	\int_x\,\bigg[\frac{1}{4}V^a_{\mu\nu}V^{a,\mu\nu}
	+(D_\mu S_i)^\dagger(D^\mu S_i)+V(\Phi,S)\bigg]\nn\\[1ex]
	&+S_{\gf}^{\SU(2)_\D}+S_{\rm ghosts}^{\SU(2)_\D},
\end{align}
where $V^a_{\mu\nu}=\pt_\mu V^a_\nu-\pt_\nu V^a_\mu+g_{\D}\epsilon^a{}_{bc}V^b_{\mu}V^c_{\nu}$ is the field-strength of the $\SU(2)_\D$ gauge field and
\begin{equation}
	D_\mu S_i=\partial_\mu S_i-i g_{\D} V_{a,\mu} T^{a}_{ij} S_j,
\end{equation}
where we use the same notation for the non-Abelian dark gauge coupling as in the Abelian case; these couplings can be distinguished by the context. The matrices $T^a$ are the generators of the $\SU(2)_\D$ dark gauge group. Gauge symmetry forbids kinetic mixing of the DM vector triplet $V^i_\mu$ with SM gauge fields.  For this non-Abelian case, Faddeev-Popov ghosts do not decouple in the matter sector. In this way, the proper gauge-fixing action along with the associated Faddeev-Popov ghost term are given by
\begin{align}
	S_{\gf}^{\SU(2)_\D}&+S_{\rm ghosts}^{\SU(2)_\D}=\frac{1}{2\xi_{\D}}\int_{\bar{x}}\left(\pt_\mu V^{a,\mu} \right)^2\nn\\[1ex]
	&+\int_{\bar{x}}\left(\bar{c}^{(\D)}_a\pt_\mu\pt^\mu c^{(\D)}_a-g_{\D}\epsilon^{abc}\bar{c}^{(\D)}_a \pt_\alpha\left(V_{c}{}^{\alpha}c^{(\D)}_b\right)\right).
\end{align}
 The Landau-gauge limit is also chosen here. Similarly to the Abelian case, the potential is chosen with the normalization
 \al{\label{eq::SU2potential}
	V(\Phi,S)&=m^2_{\H}\Phi_i^\dagger \Phi_i+\frac{\lambda_\H}{6}(\Phi_i^\dagger \Phi_i)^2+m^2_{\S} S_i^\dagger S_i\nn\\[1ex]
	&+\frac{\lambda_\S}{6}(S_i^\dagger S_i)^2+2\lambda_{p}(\Phi_i^\dagger \Phi_i)(S_i^\dagger S_i).
}

\section{Beta functions}\label{sec:betas}
\subsection{Functional renormalization group}

In the present paper, we use the functional renormalization group (FRG) \cite{Wetterich:1992yh,Morris:1993qb,Ellwanger:1993mw} as a tool to derive beta functions in gravity-matter systems. The key ingredient of the FRG is the flowing action $\Gamma_k$, a functional that describes the effective dynamics of Euclidean quantum field theories after integrating out fluctuations characterized by a momentum-scale $p=(p\cdot p)^{1/2}$ larger than the infrared cutoff $k$. 
One uses $k$ to represent a renormalization group scale that separates ultraviolet from infrared modes.
In practice, one introduces $k$ by adding a regulator function $\mathbf{R}_k$ to the microscopic action $S$, acting as a momentum-dependent mass term that suppresses fluctuations with momentum below $k$.

The flowing action $\Gamma_k$ satisfies the flow equation,
\begin{eqnarray}\label{eq:floweq}
    k \partial_k \Gamma_k = 
    \frac{1}{2} \text{STr} \left[ \left( \Gamma_k^{(2)} + \textbf{R}_k \right)^{-1} k\partial_k \textbf{R}_k\right] \,,
\end{eqnarray}
where $\Gamma_k^{(2)}$ is the 2-point function obtained by taking functional derivatives of $\Gamma_k$ with respect to fields. $\text{STr}$ is the supertrace, which is a trace over all indices and space-time coordinates with appropriate signs/multiplicities in the case of complex/fermionic fields.
For more details on the FRG and its applications in quantum gravity, see, e.g., \cite{Dupuis:2020fhh} for a recent review and \cite{Reichert:2020mja,Pawlowski:2020qer,Pawlowski:2023gym,Saueressig:2023irs} for introductions.

One important difference between beta functions computed with the FRG in comparison with perturbative renormalization group schemes (e.g., $\overline{\text{MS}}$-scheme) is that the structure $( \Gamma_k^{(2)} + \textbf{R}_k )^{-1}$ in the flow Eq. \eqref{eq:floweq} generically produces threshold contributions of the form $(1+\text{mass}^2)^{-\#}$ when one integrates over massive fields. Such threshold effects automatically account for the decoupling of massive fields at RG scales below their mass scale.
This is a significant contrast with perturbative schemes where the decoupling needs to be implemented by hand.

Despite being formally exact, the practical use of the flow Eq.~\eqref{eq:floweq} requires some form of approximation method. One typically uses truncation methods, where one starts from an ansatz $\Gamma_k$ and then  projects the flow Eq.~\eqref{eq:floweq} into a functional space that contains only the operators that were already included in the original ansatz.
In this paper, we work with a truncation for both Abelian and non-Abelian DM models, where $\Gamma_k$ has the same functional form as the classical actions \eqref{eq::actionU1} and \eqref{eq::actionSU2}, respectively, but promoting the couplings to be dependent on the renormalization group scale $k$. This truncation is motivated by previous results that asymptotically safe gravity can induce a near-perturbative UV completion for matter models, see, e.g., \cite{Eichhorn:2022gku} and references therein. In such a near-perturbative setting, the above truncation is likely to capture all relevant terms and is thus sufficient for a first study.

For practical calculations with the FRG we need to specify the regulator function $\textbf{R}_k$. In this paper, we choose
\begin{eqnarray}
    \textbf{R}_k(p^2) = \left(\Gamma^{(2)}(p)-\Gamma^{(2)}(0)\right) \,r(p^2/k^2) \,,
\end{eqnarray}
where $r(y)$ is the Litim shape function \cite{litim:2001org},
\begin{eqnarray}
    r(y) = \left( \frac{1}{y^\alpha} - 1 \right) \theta(1-y) \,,
\end{eqnarray}
with $\alpha=1$, except for spinor fields where we use $\alpha=1/2$.

\subsection{Beta functions}

Using the FRG, we computed the beta functions of SM and DM couplings including contributions due to gravitational fluctuations. In this section, we present explicit expressions for the beta function used in our analysis.

For compactness, we define the effective Newton coupling
\begin{equation}
    G_{(m_1,m_2)}^n = \frac{G^n}{(1-2\Lambda)^{m_1} (1-4\Lambda/3)^{m_2}},  
\end{equation}
where $G$ and $\Lambda$ are the dimensionless versions of the scale-dependent Newton coupling and cosmological constant, respectively. For $n=1$, we use the notation $G_{(m_1,m_2)} = G_{(m_1,m_2)}^{1}$.
Some of the beta functions also depend on a parameter $\zeta$, which allows us to unify the results obtained with $\text{U(1)}_\text{D}$ ($\zeta = 0$) and $\text{SU(2)}_\text{D}$ ($\zeta=1$) vector DM models.

For the SM (non-)Abelian gauge couplings, we find 
\begin{equation}\label{eq:betagY}
    \beta_{g_\text{Y}} = 
    - f_g \, g_\text{Y} +
    \frac{5\,g_\text{Y}^3}{12 \pi ^2} +
    \frac{g_\text{Y}^3}{96 \pi ^2 (1 + m_\text{H}^2)^4} \,,
\end{equation}
\begin{equation}\label{eq:betag2}
    \beta_{g_2} =
    - f_g \, g_\text{2}
    -\frac{5\,g_2^3}{24 \pi ^2} +
    \frac{g_2^3}{96 \pi ^2 (1 + m_\text{H}^2)^4}  \,,
\end{equation}
\begin{equation}\label{eq:betag3}
    \beta_{g_3} =
    - f_g \, g_\text{3} 
    -\frac{7\,g_3^3}{16 \pi ^2}\,,
\end{equation}
where $f_g$ is the gravitational contribution to the flow of gauge couplings, see \cite{Daum:2009dn,Daum:2010bc,Folkerts:2011jz,Harst:2011zx,Christiansen:2017cxa,Christiansen:2017gtg,Eichhorn:2017lry,Eichhorn:2019yzm,DeBrito:2019gdd,Eichhorn:2021qet,deBrito:2022vbr}
\begin{equation}
    f_g = \frac{5 \,G_{(1,0)}}{9 \pi}-\frac{5 \,G_{(2,0)}}{18 \pi} \,.
\end{equation}
We note that $f_g$ is the same for all gauge couplings, which is a consequence of gravity being ``blind'' to internal symmetries.

For the gauge coupling in the dark sector, we find
\begin{equation}\label{eq:betagD}
    \beta_{g_\text{D}} = - f_g \,g_\text{D} 
    + \frac{(1-\zeta/2)\,g_\text{D}^3}{48 \pi ^2 (1 + m_\S^2)^4} -\frac{11 \,\zeta\,g_\text{D}^3}{24\pi^2} \,.
\end{equation}

In the Yukawa sector of the SM, we focus on the top and bottom Yukawa couplings.
The other flavors have subleading effects in  our analysis.
The resulting beta functions are
\begin{equation}\label{eq:betayt}
    \begin{aligned}
        \beta_{y_t} &= 
        - f_y \,y_t + \frac{3\,y_t^3}{16 \pi ^2}
        +\frac{3\,y_t^3}{32 \pi ^2 (1+m_\text{H}^2)^2} \\
        &+\frac{3\,y_b^2 y_t}{16 \pi ^2}
        -\frac{y_b^2 y_t}{16 \pi ^2 (1+m_\text{H}^2)} 
        -\frac{y_b^2 y_t}{32 \pi ^2 (1+m_\text{H}^2)^2} \\
        &-\frac{g_\text{Y}^2 \,y_t}{24 \pi ^2}
        -\frac{3\,g_\text{Y}^2 \,y_t}{128 \pi ^2 (1+m_\text{H}^2)}
        - \frac{3\,g_\text{Y}^2 \,y_t}{128 \pi ^2 (1+m_\text{H}^2)^2} \\
        & -\frac{9\,g_2^2\,y_t}{128 \pi ^2 (1+m_\text{H}^2)}
        - \frac{9\,g_2^2\,y_t}{128 \pi ^2 (1+m_\text{H}^2)^2} 
        -\frac{g_3^2\,y_t}{2 \pi ^2} \,,
    \end{aligned}
\end{equation}
and
\begin{equation}\label{eq:betayb}
    \begin{aligned}
        \beta_{y_b} &= 
        - f_y \,y_b + \frac{3\,y_b^3}{16 \pi ^2}
        +\frac{3\,y_b^3}{32 \pi ^2 (1+m_\text{H}^2)^2} \\
        &+\frac{3\,y_t^2 y_b}{16 \pi ^2}
        -\frac{y_t^2 y_b}{16 \pi ^2 (1+m_\text{H}^2)} 
        -\frac{y_t^2 y_b}{32 \pi ^2 (1+m_\text{H}^2)^2} \\
        &+\frac{g_\text{Y}^2 \,y_b}{48 \pi ^2}
        -\frac{3\,g_\text{Y}^2 \,y_b}{128 \pi ^2 (1+m_\text{H}^2)}
        - \frac{3\,g_\text{Y}^2 \,y_b}{128 \pi ^2 (1+m_\text{H}^2)^2} \\
        & -\frac{9\,g_2^2\,y_b}{128 \pi ^2 (1+m_\text{H}^2)}
        - \frac{9\,g_2^2\,y_b}{128 \pi ^2 (1+m_\text{H}^2)^2} 
        -\frac{g_3^2\,y_b}{2 \pi ^2} \,,
    \end{aligned}
\end{equation}
with the flavor-independent gravitational contribution, see \cite{Oda:2015sma,Eichhorn:2016esv,Eichhorn:2017eht,DeBrito:2019gdd,Eichhorn:2020sbo,deBrito:2022vbr}
\begin{equation}
    \begin{aligned}
        f_y &= 
        -\frac{15\,G_{(2,0)}}{16 \pi} 
        +\frac{G_{(0,1)}}{8 \pi}
        -\frac{G_{(0,2)}}{48 \pi} \\
        &+\frac{2\,G_{(0,2)}\,m_\text{H}^4}{3\pi (1\!+\!m_\text{H}^2)^2}
        -\frac{2\,G_{(0,1)}\,m_\text{H}^2}{5\pi (1\!+\!m_\text{H}^2)}
        +\frac{8\,G_{(0,1)}\,m_\text{H}^2}{15\pi (1\!+\!m_\text{H}^2)^2}  \\
        &-\frac{2\,G_{(0,2)}\,m_\text{H}^2}{15\pi (1\!+\!m_\text{H}^2)}
        -\frac{G_{(0,1)}}{36\pi (1\!+\!m_\text{H}^2)^2} 
        -\frac{G_{(0,2)}}{36\pi (1\!+\!m_\text{H}^2)} \,. 
    \end{aligned}
\end{equation}

For the Higgs and dark quartic scalar couplings, we find
\begin{equation}\label{eq:betalambdaH}
    \begin{aligned}
        \beta_{\lambda_\text{H}} &= 
        - f_{\lambda_\text{H}} \, \lambda_\text{H}
        +\frac{\lambda_\text{H}^2}{4 \pi ^2 (1+m_\text{H}^2)^3} \\
        &-\frac{9\,g_2^2\,\lambda_\text{H}}{32 \pi ^2 (1+m_\text{H}^2)} 
        -\frac{9\,g_2^2\,\lambda_\text{H}}{32 \pi ^2 (1+m_\text{H}^2)^2} \\
        &-\frac{3\,g_\text{Y}^2\,\lambda_\text{H}}{32 \pi ^2 (1+m_\text{H}^2)}
        -\frac{3\,g_\text{Y}^2\,\lambda_\text{H}}{32 \pi ^2 (1+m_\text{H}^2)^2} \\
        &+\frac{3\,(1+\zeta)\,\lambda_\text{p}^2}{2 \pi ^2 (1+m_\S^2)^3}
        +\frac{3\,\lambda_\text{H}\,y_b^2}{4 \pi ^2}
        +\frac{3\,\lambda_\text{H}\,y_t^2}{4 \pi ^2} \\
        &+\frac{9\,g_2^2\,g_\text{Y}^2}{32 \pi ^2}
        +\frac{27\,g_2^4}{64\pi^2}+\frac{9 g_\text{Y}^4}{64 \pi ^2}
        -\frac{9\,y_b^4}{4\pi^2}
        -\frac{9\,y_t^4}{4\pi^2} \,,
\end{aligned}
\end{equation}
and
\begin{equation}\label{eq:betalambdaS}
    \begin{aligned}
        \beta_{\lambda_\S} &= 
        -f_{\lambda_\S} \, \lambda_\S
        +\frac{5\,(1+\zeta/5)\,\lambda_\S^2}{24 \pi ^2 (1+m_\S^2)^3}
        +\frac{3\,\lambda_\text{p}^2}{\pi^2 (1+m_\text{H}^2)^3} \\
        &-\frac{3\,(1-\zeta/4)\,g_\text{D}^2\,\lambda_\S }{8 \pi ^2 (1+m_\S^2)^2}
        -\frac{3\,(1-\zeta/4)\,g_\text{D}^2 \,\lambda_\S}{8 \pi ^2 (1+m_\S^2)} \\
        &+\frac{9\,(1-13\,\zeta/16)\,g_\text{D}^4}{4 \pi ^2} \,.
    \end{aligned}
\end{equation}
The gravitational contributions to the quartic scalar couplings are, see, e.g., \cite{Narain:2009fy,Eichhorn:2017als,Pawlowski:2018ixd, Eichhorn:2020sbo,Wetterich:2019rsn} 
\begin{align}
    f_{\lambda_\text{H/S}} &= - \frac{5\,G_{(2,0)}}{2 \pi}
    -\frac{G_{(0,2)}}{3\pi} + F_{\lambda_\text{H/S}}(m_\text{H/S}^2) \,,
\end{align}
with
\begin{equation}
    \begin{aligned}
        F_\lambda(m^2) &=
        -\frac{16\,G_{(0,1)}\,m^4}{\pi(1+m^2)^3} 
        -\frac{16\,G_{(0,2)}\,m^4}{3\pi(1+m^2)^2}  \\
        &+\frac{16\,G_{(0,1)}\,m^2}{3\pi(1+m^2)^2} 
        +\frac{8\,G_{(0,2)}\,m^2}{3\pi(1+m^2)} \\
        &-\frac{G_{(0,1)}}{9\pi(1+m^2)^2} 
        -\frac{G_{(0,2)}}{9\pi(1+m^2)} \\
        &+\lambda^{-1}\,\bigg( 
        \frac{64\,m^4}{3} G_{(0,3)}^2 + 240\,m^4\,G_{(3,0)}^2 \\
        &\quad -\frac{256\,m^6\,G_{(0,2)}^2}{3 \left(1+m^2\right)^2}
        +\frac{1024\, m^8\,G_{(0,2)}^2}{3 \left(1+m^2\right)^3}\\
        &\quad +\frac{1024\,m^8\,G_{(0,3)}^2}{3 \left(1+m^2\right)^2}
        -\frac{512\,m^6\,G_{(0,3)}^2}{3 \left(1+m^2\right)}
        \bigg)
        \,.
    \end{aligned}
\end{equation}
As in the gravitational contribution to the flow of the SM gauge couplings, we note that these contributions are the same for both quartic couplings, \ie, the gravitational contribution is ``blind" to internal symmetries.
For the scalar portal coupling, we find
\begin{equation}\label{eq:betalambdap}
    \begin{aligned}
    \beta_{\lambda_\text{p}} &= - f_{\lambda_\text{p}} \, \lambda_\text{p} 
    +\frac{\lambda_\text{p}^2}{4 \pi ^2 (1+m_\text{H}^2)^2 (1+m_\S^2)} \\
    &+\frac{\lambda_\text{p}^2}{4 \pi ^2 (1+m_\text{H}^2) (1+m_\S^2)^2}
    -\frac{9\,g_2^2\,\lambda_\text{p}}{64 \pi ^2 (1+m_\text{H}^2)} \\
    &-\frac{9\,g_2^2\,\lambda_\text{p}}{64 \pi ^2 (1+m_\text{H}^2)^2}
    -\frac{3\,(1-\zeta/4)\,g_\text{D}^2\,\lambda_\text{p}}{16 \pi ^2 (1+m_\S^2)} \\
    &-\frac{3\,(1-\zeta/4)\,g_\text{D}^2\,\lambda_\text{p}}{16 \pi ^2 (1+m_\S^2)^2} 
    -\frac{3\,g_\text{Y}^2\,\lambda_\text{p}}{64 \pi ^2 (1+m_\text{H}^2)} \\
    &-\frac{3\,g_\text{Y}^2\,\lambda_\text{p}}{64 \pi ^2(1+m_\text{H}^2)^2}
    +\frac{\lambda_\text{H}\,\lambda_\text{p}}{8 \pi ^2 (1+m_\text{H}^2)^3} \\
    &+\frac{(1+\zeta/2)\,\lambda_\text{p}\,\lambda_\S}{12 \pi ^2 (1+m_\S^2)^3}
    +\frac{3\,\lambda_\text{p}\,y_b^2}{8 \pi^2}
    +\frac{3\,\lambda_\text{p}\,y_t^2}{8 \pi ^2} \,,
    \end{aligned}
\end{equation}
with gravitational contribution (see \cite{Eichhorn:2017als, Eichhorn:2020sbo})
\begingroup
\allowdisplaybreaks
    \begin{align}
        &f_{\lambda_\text{p}} =
        -\frac{5\,G_{(2,0)}}{2\pi} 
        -\frac{G_{(0,2)}}{3\pi} 
        -\frac{8\,G_{(0,1)}\,m_\text{H}^4}{3\pi(1+m_\text{H}^2)^3} \nonumber \\
        &-\frac{16\,G_{(0,1)}\,m_\text{H}^2 m_\S^2}{3\pi(1+m_\text{H}^2)^2(1+m_\S^2)} 
        -\frac{16\,G_{(0,1)}\,m_\text{H}^2 m_\S^2}{3\pi(1+m_\text{H}^2)(1+m_\S^2)^2} \nonumber  \\
        &+\frac{4\,G_{(0,2)}\,m_\text{H}^2}{3\pi(1+m_\text{H}^2)} 
        -\frac{G_{(0,1)}}{18 \pi(1+m_\text{H}^2)^2} 
        -\frac{G_{(0,2)}}{18\pi(1+m_\text{H}^2)} \nonumber \\
        &-\frac{8\,G_{(0,1)}\,m_\S^4}{3\pi(1+m_\S^2)^3}
        +\frac{8\,G_{(0,1)}\, m_\S^2}{3 \pi (1+m_\S^2)^2} 
        -\frac{G_{(0,1)}}{18\pi(1+m_\S^2)^2}\\
        &-\frac{16\,G_{(0,2)}\,m_\text{H}^2 m_\S^2}{3 \pi (1+m_\text{H}^2) (1+m_\S^2)} 
        +\frac{8\,G_{(0,1)}\,m_\text{H}^2}{3\pi(1+m_\text{H}^2)^2} \nonumber \\
        &+\frac{4\,G_{(0,2)}\,m_\S^2}{3\pi(1+m_\S^2)} 
        -\frac{G_{(0,2)}}{18\pi(1+m_\S^2)}\nonumber  \\
        &+\lambda_\text{p}^{-1} \bigg( 
        +\frac{32\, G_{(0,3)}^2 \,m_\text{H}^2 \,m_{\S }^2}{9} + 40\,G_{(3,0)}^2\,m_\text{H}^2\,m_{\S }^2 \nonumber \\
        &\quad-\frac{128\,G_{(0,3)}^2\,m_\text{H}^4\,m_{\S }^2}{9 \left(1+m_\text{H}^2\right)}
        -\frac{64\,G_{(0,2)}^2\,m_\text{H}^4\,m_{\S}^2}{9 \left(1+m_\text{H}^2\right)^2} \nonumber \\
        &\quad+\frac{512\,G_{(0,3)}^2\,m_\text{H}^4\,m_{\S}^4}{9 \left(1+m_\text{H}^2\right) \left(1+m_{\S}^2\right)}
        +\frac{256\,G_{(0,2)}^2\,m_\text{H}^4\,m_{\S}^4}{9 \left(1+m_\text{H}^2\right)^2 \left(1+m_{\S}^2\right)} \nonumber \\
        &\quad+\frac{256\,G_{(0,2)}^2\,m_\text{H}^4 m_{\S}^4}{9 \left(1+m_\text{H}^2\right)\left(1+m_{\S }^2\right)^2}
        -\frac{128\,G_{(0,3)}^2\,m_\text{H}^2\,m_{\S }^4}{9 \left(1+m_{\S}^2\right)} \nonumber \\
        &\quad-\frac{64\,G_{(0,2)}^2\,m_\text{H}^2\,m_{\S }^4}{9 \left(1+m_{\S}^2\right)^2}\bigg)
        \,. \nonumber 
    \end{align}
\endgroup

There is again the same universality, \ie, those terms that are independent of the scalar masses are exactly the same as the first two terms in $f_{\lambda_{\rm H/S}}$.

Finally, by defining dimensionless versions of the Higgs and dark mass parameters by the  rescalings $m_\H^2\rightarrow k^2 m_\H^2$ and $m_\S^2\rightarrow k^2 m_\S^2$, their beta functions read
\begin{equation}\label{eq:betamH2}
    \begin{aligned}
        \beta_{m_\text{H}^2} &= -2 m_\text{H}^2 - f_{m_\text{H}^2} \, m_\text{H}^2
        -\frac{9 g_2^2 m_\text{H}^2}{64 \pi^2 (1+m_\text{H}^2)} \\
        &-\frac{9 g_2^2 m_\text{H}^2}{64 \pi^2 (1+m_\text{H}^2)^2}
        -\frac{3 g_\text{Y}^2 m_\text{H}^2}{64 \pi^2 (1+m_\text{H}^2)} \\
        &-\frac{3 g_\text{Y}^2 m_\text{H}^2}{64\pi^2 (1+m_\text{H}^2)^2} 
        +\frac{3 m_\text{H}^2 y_b^2}{8 \pi^2}
        +\frac{3 m_\text{H}^2 y_t^2}{8 \pi^2} \\
        &-\frac{9 g_2^2}{64 \pi^2}
        -\frac{3 g_\text{Y}^2}{64 \pi^2}
        -\frac{\lambda_\text{H}}{16 \pi^2 (1+m_\text{H}^2)^2} \\
        &-\frac{(1+\zeta)\,\lambda_\text{p}}{8 \pi^2 (1+m_\S^2)^2}
        +\frac{3 y_b^2}{8 \pi^2}
        +\frac{3 y_t^2}{8 \pi^2} \,,
    \end{aligned}
\end{equation}
and
\begin{equation}\label{eq:betamS2}
    \begin{aligned}
        \beta_{m_\S^2} &= - 2 m_\S^2 - f_{m_\S^2} \, m_\S^2
        - \frac{(1+\zeta/2)\,\lambda_\S }{24 \pi^2 (1 + m_\S^2)^2}\\
        &-\frac{3\,(1-\zeta/4)\,g_\text{D}^2 m_\S^2}{16 \pi^2 (1 + m_\S^2)^2}
        -\frac{3\,(1-\zeta/4)\,g_\text{D}^2 m_\S^2}{16 \pi^2 (1 + m_\S^2)} \\
        &-\frac{3\,(1-\zeta/4)\,g_\text{D}^2}{16 \pi^2} 
        -\frac{\lambda_\text{p}}{4 \pi^2 (1+m_\text{H}^2)^2}\,,
    \end{aligned}
\end{equation}
with gravitational contribution
\begin{equation}
    \begin{aligned}
        f_{m^2} &= 
        -\frac{5\,G_{(2,0)}}{2\pi}
        -\frac{G_{(0,2)}}{3\pi} 
        +\frac{4\,G_{(0,2)}\,m^4}{3 \pi  \left(1+m^2\right)^2} \\
        &+\frac{4\,G_{(0,1)}\,m^2}{3 \pi  \left(1+m^2\right)^2} 
        -\frac{G_{(0,1)}}{18 \pi  \left(1+m^2\right)^2} \\
        &-\frac{G_{(0,2)}}{18 \pi  \left(1+m^2\right)} \,
    \end{aligned}
\end{equation}
for $m^2 = m_\text{H}^2$ and $m^2 = m_\S^2$.

Besides the gravitational contributions, the results presented here differ from standard one-loop beta functions in two ways: i) they automatically contain threshold effects due to a masslike regulator function; ii) the beta functions for the mass parameters contain terms proportional to $\lambda_\text{H}$, $\lambda_\S$, and $\lambda_\text{p}$ that are nonvanishing in the limit $m_\text{H}^2,m_\S^2 \to 0$.

In our analysis of this system of beta functions, we have considered the following approximations:
\begin{itemize}
    \item We set $m_\text{H}^2$ and $m_\S^2$ to zero in the beta functions of the gauge and Yukawa couplings. This approximation allows us to integrate the flow of the gauge and Yukawa couplings independently of the quartic couplings and mass parameters.
    \item We parametrize the flow of the gravitational couplings $G$ and $\Lambda$ with a Heaviside function according to
\begin{align}
    G(k) = G_* \,\theta(k - M_\text{Pl}) \,
    \,\, \text{and}\,\,
    \Lambda(k) = \Lambda_* \, \theta(k - M_\text{Pl})\,,
\end{align}
\end{itemize}
where $G_{\ast}$ and $\Lambda_{\ast}$ denote their fixed-point values, corresponding to the zero of the gravitational beta functions in \cite{Eichhorn:2017ylw}. This parametrization is a good approximation of the flow resulting from the integration of the beta functions in \cite{Eichhorn:2017ylw}. In particular, it implements the decoupling of the gravitational contributions below the Planck scale. 
We note that this decoupling is not due to true massive threshold effects, but due to the fact that the (dimensionful) Newton coupling is approximately constant below the Planck scale, \ie, it is due to gravity transitioning into the classical-gravity regime.

\section{Results}\label{sec:results}
\subsection{Constraints from direct detection experiments}\label{Subsub::direct_detection}

 There are several ways of constraining DM models in the IR including via direct and indirect detection, hidden decays and from requiring the correct relic density. We assume that the presented models account for all of dark matter. We, thus, require a density of $\Omega_c h^2 = 0.120 \pm 0.001$ \cite{Planck:2018vyg}. This requirement is satisfied on a single line in the coupling-mass parameter space.

An overview of the constraints of the U(1)$_\D$ model can be found in \cite{Frandsen:2022klh}, which focuses on DM masses $M_V < \mathcal{O}(10)$ TeV. Then, the DM mass is constrained to around 1 TeV with a coupling $0.66 \le g_{\D} \le 0.7$. 

 In the following section we will investigate three points in parameter space to illustrate the general situation. Besides one point in the phenomenologically viable area, we also investigate two points at lower values of the coupling, $g_{\D}=0.1$ and $g_{\D}=0.25$. Despite being  excluded due to an overproduction of the DM relic density from thermal freeze-out according to \cite{Frandsen:2022klh}, we include them in our study to illustrate the exclusion mechanism from asymptotic safety. Furthermore, assuming a mass of $M_V=911\, \textmd{GeV}$, these parameter points are also excluded by direct detection experiments such as XENON1T and LZ \cite{xenon:2017, aalbers:2023fdm}.

For the SU(2)$_\D$ model, there is a small unconstrained region similar to the U(1)$_\D$ model for $0.7 \le g_{\D} \le 0.8$. 
Alternatively, one can consider the case where the coupling is $g_{\D}\geq 2$, where the model generally escapes constraints from direct detection. Arguably this comes at the price of departing from the perturbative regime; depending on how heavy a DM mass one considers, the larger the mass, the larger the coupling \cite{Frandsen:2022klh}.

\subsection{Constraints from asymptotic safety on $\U(1)_\D$ dark matter}\label{Sec::U1_case}
In this section, we explore asymptotically safe RG trajectories emanating from UV fixed points obtained from the beta functions presented in the previous section. Here, we focus on the $\text{U}(1)_\text{D}$ DM model to discover whether asymptotically safe constraints are compatible with the phenomenological constraints reviewed above.

We focus on fixed-point solutions with vanishing non-Abelian gauge and bottom Yukawa couplings, \ie,
\begin{equation}\label{eq:g2g3yb}
    g_{2,\,\ast}= 0\,,\quad g_{3,\,\ast}= 0 \quad\text{and}\quad y_{b,\,\ast}=0 \,.  
\end{equation}
This choice is motivated by the following: i) the fact that $g_2$ and $g_3$ are asymptotically free even without gravity; ii) previous studies in asymptotically safe quantum gravity indicating that one can accommodate the IR-measured value of $y_b$ on an RG trajectory starting from $y_{b,\,\ast}=0$ \cite{Eichhorn:2017ylw}, whereas a nonzero fixed-point value of $y_b$ results in a prediction of the ratio of the top and bottom mass \cite{Eichhorn:2018whv} that is not our focus here.

Using the beta functions reported in \cite{Eichhorn:2017ylw} to compute the fixed-point values of the gravitational couplings, we find
\begin{equation}\label{eq:FPgravU1}
    G_{\ast} = 2.78 \quad \text{and} \quad \Lambda_{\ast} = -3.58 \,,
\end{equation}
which are obtained by setting the matter content to that of the SM plus the $\text{U}(1)_\text{D}$ DM model.

In total, there are eight fixed-point (FP) candidates compatible with Eq.~\eqref{eq:g2g3yb} and Eq.~\eqref{eq:FPgravU1}. In Table.~\ref{table:fixed-points}, we classify the different fixed points in terms of the signs and (ir)relevance of gauge and Yukawa couplings.

First, we note that the fixed points with nonvanishing gauge coupling $g_\text{D}$ (FP's (v)-(viii), cf.~Table.~\ref{table:fixed-points}) lead to a negative (unstable) scalar potential in the fixed-point regime, \ie, at least the dark quartic scalar coupling $\lambda_\text{S}$ is negative at the fixed point.
This is a consequence of the signs of the $g_\text{D}^4$ and $g_\text{D}^2$ contributions to $\beta_{\lambda_\text{S}}$ and $\beta_{m_\text{S}^2}$, respectively, cf.~Eq.~\eqref{eq:betalambdaS} and Eq.~\eqref{eq:betamS2}. That these terms can trigger negative quartic couplings is a generic result and holds whenever the gravitational contribution is not too large, see Table.~1 in \cite{Eichhorn:2019dhg}.

Similarly, the fixed point with $g_{{\rm Y},\,\ast}>0$ and $y_{t,\,\ast}=0$ (FP (iv), cf.~Table.~\ref{table:fixed-points}) leads to negative fixed-point values for $\lambda_\text{H}$ and $m_\text{H}^2$ through a similar type of terms in the respective beta functions. 

In our study, we stay within a near-perturbative regime and therefore assume that the quadratic and quartic terms in the potential are sufficient. We further assume that the fixed-point potential should be stable in order to yield a well-defined path integral. Thus, we will not investigate RG trajectories emanating from fixed-point candidates with negative values of quartic couplings.

Second, we also disregard the fixed point (ii) in Table.~\ref{table:fixed-points}. This is motivated by previous studies in asymptotically safe quantum gravity indicating that a fixed point with $g_{{\rm Y},\,\ast}>0$ leads to a prediction of the IR value of the hypercharge gauge coupling that is approximately 35\% larger than the measured value \cite{Eichhorn:2017lry}. The fixed-point value here is different than in \cite{Eichhorn:2017lry}, because it is affected by the dark degrees of freedom which change the gravitational fixed-point values and therefore the fixed-point values of matter couplings. Nevertheless, the change is not sufficient to produce the correct value of the gauge coupling in the IR. The fixed-point value of the hypercharge gauge coupling with DM degrees of freedom is $g_{{\rm Y},\,\ast}=1.143$, whereas its Planck-scale value in the Standard Model is significantly smaller.

\begin{table*}[hbt!]
\begin{tabular}{ |p{1.0cm} ||p{1.3cm}||p{1.3cm}||p{1.3cm}|p{1.3cm}||p{1.3cm}||p{1.3cm}||p{2.6cm}|}
\hline
\textbf{FP} & $g_{{\rm D},\,\ast}$ & $g_{{\rm Y},\,\ast}$  & $g_{2,\,\ast}$ & $g_{3,\,\ast}$ & $y_{t,\,\ast}$ & $y_{b,\,\ast}$ & Stable $V_*(\Phi,S)$? \\ \hline\hline
i           & 0, rel       & 0, rel      & 0, rel    & 0, rel  & 0, rel      & 0, rel  & \hspace{.2pt} Yes  (Flat)\\ \hline
ii          & 0, rel       & $>0$, irrel & 0, rel    & 0, rel  & $>0$, irrel & 0, rel  & \hspace{.2pt} Yes  \\ \hline
iii         & 0, rel       & 0, rel      & 0, rel    & 0, rel  & $>0$, irrel & 0, rel  & \hspace{.2pt} Yes  \\ \hline
iv          & 0, rel       & $>0$, irrel & 0, rel    & 0, rel  & 0, rel      & 0, rel  & \hspace{.2pt} No   \\ \hline
v           & $>0$, irrel  & 0, rel      & 0, rel    & 0, rel  & 0, rel      & 0, rel  & \hspace{.2pt} No   \\ \hline
vi          & $>0$, irrel  & $>0$, irrel & 0, rel    & 0, rel  & 0, rel      & 0, rel  & \hspace{.2pt} No   \\ \hline
vii         & $>0$, irrel  & 0, rel      & 0, rel    & 0, rel  & $>0$, irrel & 0, rel  & \hspace{.2pt} No   \\ \hline
viii        & $>0$, irrel  & $>0$, irrel & 0, rel    & 0, rel  & $>0$, irrel & 0, rel  & \hspace{.2pt} No   \\ \hline
\end{tabular}
\caption{Classification of fixed-point solutions for the gauge-Higgs-top-bottom system with $\U(1)_\D$ DM sector with gravity. Herein, ``rel'' stands for relevant directions and ``irrel'' denotes irrelevant directions.
$V_*(\Phi,S)$ stands for the scalar potential evaluated at the fixed-point values of the mass parameters and quartic couplings. We note that the fixed point (i) has a flat scalar potential, \ie~$V_*(\Phi,S)=0$
}
\label{table:fixed-points}
\end{table*}

The remaining viable fixed points are the candidates (i) and (iii) in Table.~\ref{table:fixed-points}. We focus on the fixed point (iii), which is the most predictive among the viable options. In particular, we note that the Yukawa coupling corresponds to an irrelevant direction at the fixed point (iii), which means that one can predict its IR value as a function of the relevant couplings. At the same time, this prediction corresponds to an upper bound on values of the Yukawa coupling achievable from fixed point (i). Due to this property, we will see that by ruling out fixed point (iii) we simultaneously rule out fixed point (i).

The quartic couplings $\lambda_\text{H}$, $\lambda_\text{S}$ and $\lambda_\text{p}$ are also irrelevant at this fixed point. Thus, their IR values are predictions of the RG trajectories emanating from such a fixed point.
Explicitly, the fixed point (iii) is located at
\begin{equation} \label{u1:fixed}
    \begin{aligned}
	&g_{{\rm Y},\,\ast}=0 \,,\qquad g_{2,\,\ast}=0 \,,\qquad g_{3,\,\ast}=0 \,,\qquad g_{{\rm D},\,\ast}=0  ,\\[1ex]
	&y_{t,\,\ast}=0.20\,, \qquad  y_{b,\,\ast}=0,\\[1ex]
	&\lambda_{{\rm H},\,\ast}=5.56\,\cdot \,10^{-3}, \qquad \lambda_{{\rm S},\,\ast}=0, \qquad \lambda_{{\rm p},\,\ast}=0,  \\[1ex]
        &(m_\H^2)_*=7.59\,\cdot \,10^{-4}, \qquad (m_\S^2)_*=0 \,.
\end{aligned}
\end{equation}
Note that the scalar potential at this fixed point is flat in the dark scalar direction.

We obtain RG trajectories by integrating the system of beta functions reported in the previous section with boundary conditions corresponding to the fixed point in \eqref{u1:fixed}.
In the integration process, we select the RG trajectories to match the appropriate IR values for relevant couplings $g_\text{Y}$, $g_2$, $g_3$ and $y_b$. In particular, we use their reference one-loop  values at $k_\text{IR} = 173\, \text{GeV}$ \cite{Buttazzo:2013uya}. We decouple the bottom Yukawa coupling by approximating $y_b \approx 0$.
In Table.~\ref{table:infrared-values}, we summarize the IR values of the various couplings at $173\,\text{GeV}$.

We investigate the three benchmark values for the $\U(1)_\D$ dark gauge coupling,
\begin{equation}
    g_{\D}(M_V=911~\textmd{GeV}) = 0.10,\, 0.25,\, 0.66 \,,
\end{equation}
fixed at a phenomenologically relevant mass scale. The latter matches the phenomenologically viable point explored in \cite{Frandsen:2022klh}.

\begin{table}
	\centering
	\begin{tabular}{ |p{2cm}||p{2cm}|p{2cm}|p{2cm}|  }
		\hline
		\multicolumn{4}{|c|}{IR values at $k=173$ GeV} \\
		\hline
		Coupling & $g_{\D}=0.10$ & $g_{\D}=0.25$ & $g_{\D}=0.66$ \\
		\hline
		$g_\Y$ (free par.)	& 0.35	& 0.35	& 0.35	\\
		$g_2$ (free par.)	& 0.65	& 0.65	& 0.65	\\
		$g_3$ (free par.)	& 1.17	& 1.17	& 1.17	\\
		$y_t$ (pred.)	& 0.665	& 0.664	& 0.665	\\
		$\lambda_\H$ (pred.)	& $-3.88\cdot 10^{-2}$ & $-4.15\cdot 10^{-2}$ & $-3.87\cdot 10^{-2}$	\\
		$\lambda_\S$ (pred.)	& $-9.77\cdot 10^{-4}$ & $-4.27\cdot 10^{-2}$	& $-4.63 \cdot 10^7$	\\
		$\lambda_{p}$ (pred.)	& $-3.13 \cdot 10^{-7}$	& $-2.16 \cdot 10^{-6}$	& -1.86	\\
		$m_\H^2$ (free par.)	& -0.005	& -0.005	& -0.005	\\
		$m_\S^2$ (free par.)	& -0.5	& -0.5	& -0.5	\\
		\hline
	\end{tabular}
	\caption{We show IR values of various couplings in the $\U(1)_\D$ DM model evaluated at $ k=173 $~GeV for different IR values of $g_{\D}$.
    The label ``free par.'' indicates that the RG flow allows us to choose the IR value of the corresponding coupling, since it is a free parameter. In contrast, the label ``pred.'' indicates that the IR value is a prediction of an RG trajectory with asymptotically safe boundary condition.
    \label{table:infrared-values}}
\end{table}

\begin{figure*}[t]
	\hspace*{-0.8cm}
	\quad \quad \ \includegraphics[width=.315\linewidth] {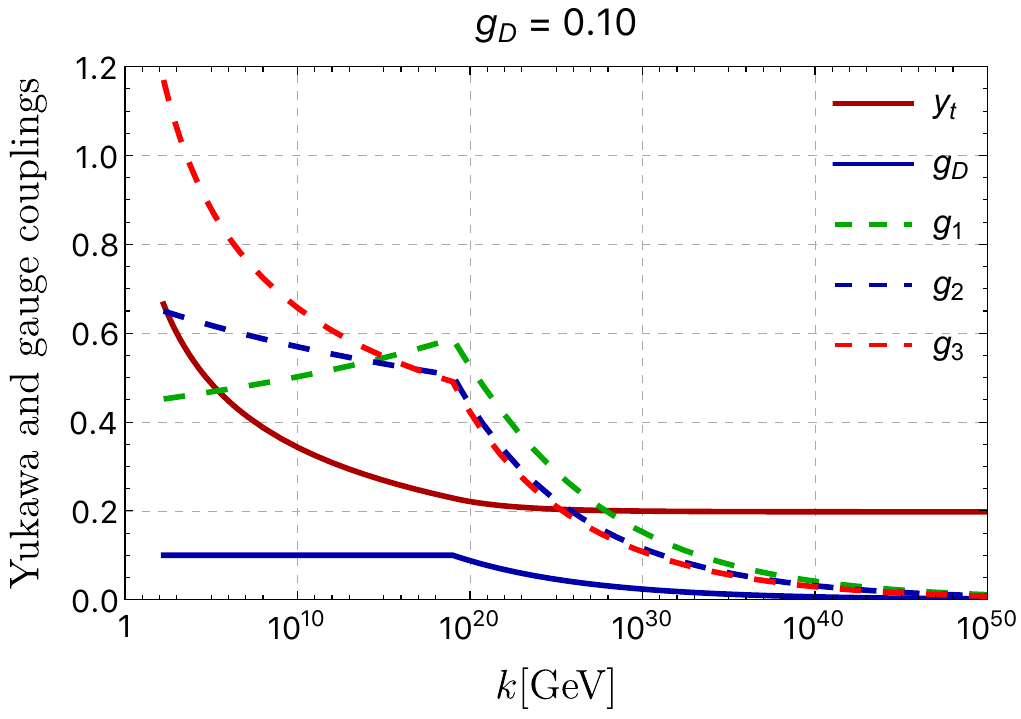} 
	\ \includegraphics[width=.315\linewidth] {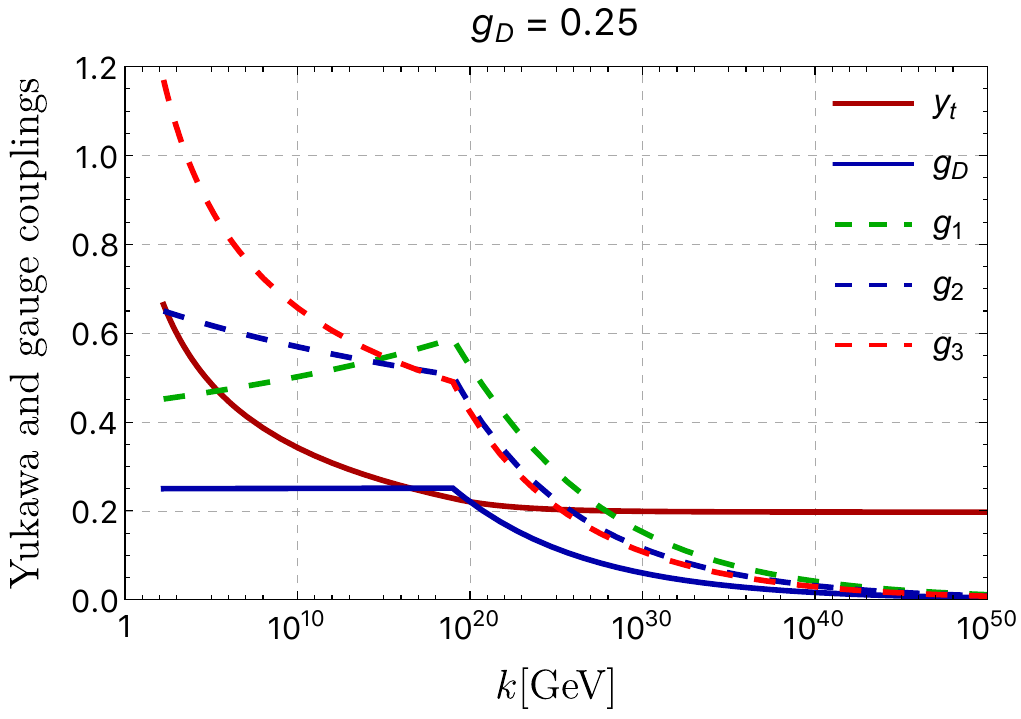}	
	\ \includegraphics[width=.315\linewidth] {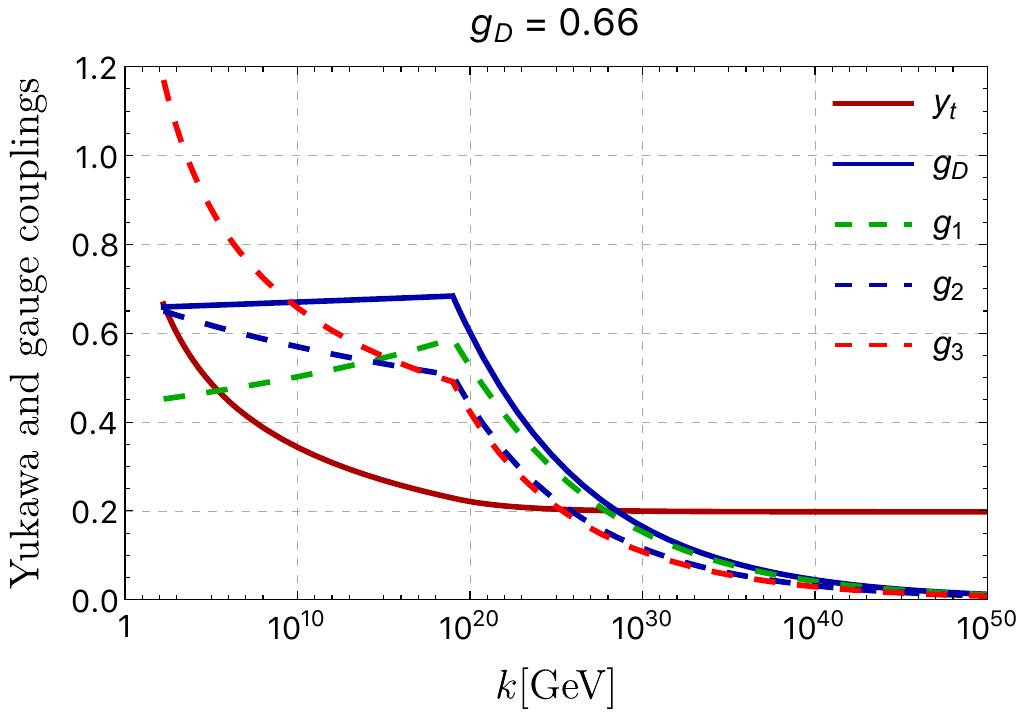}	
	\includegraphics[width=.325\linewidth] {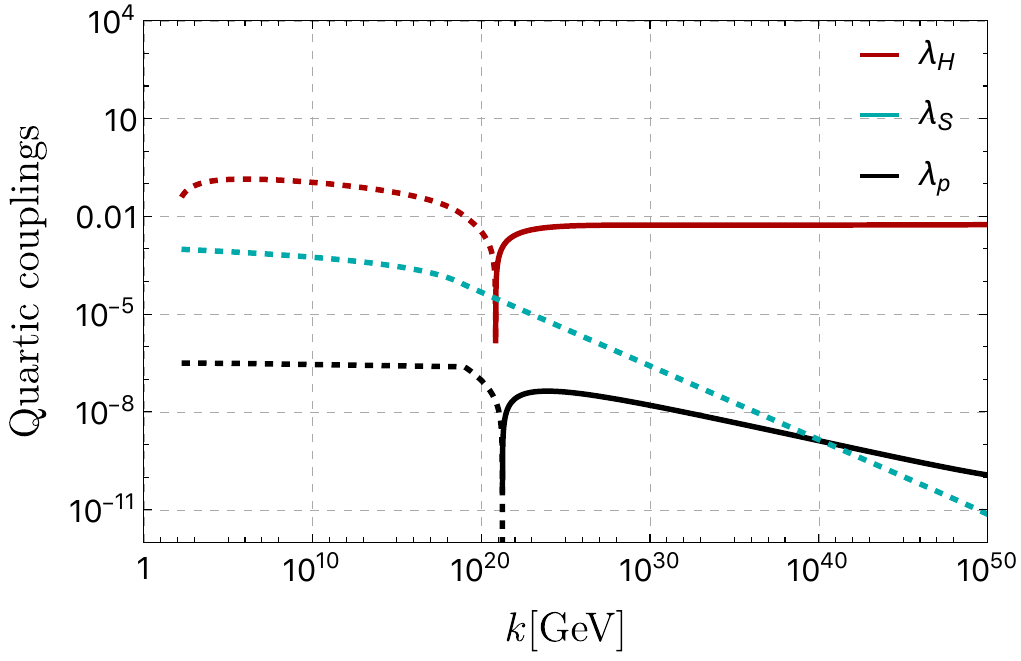}
	\includegraphics[width=.325\linewidth] {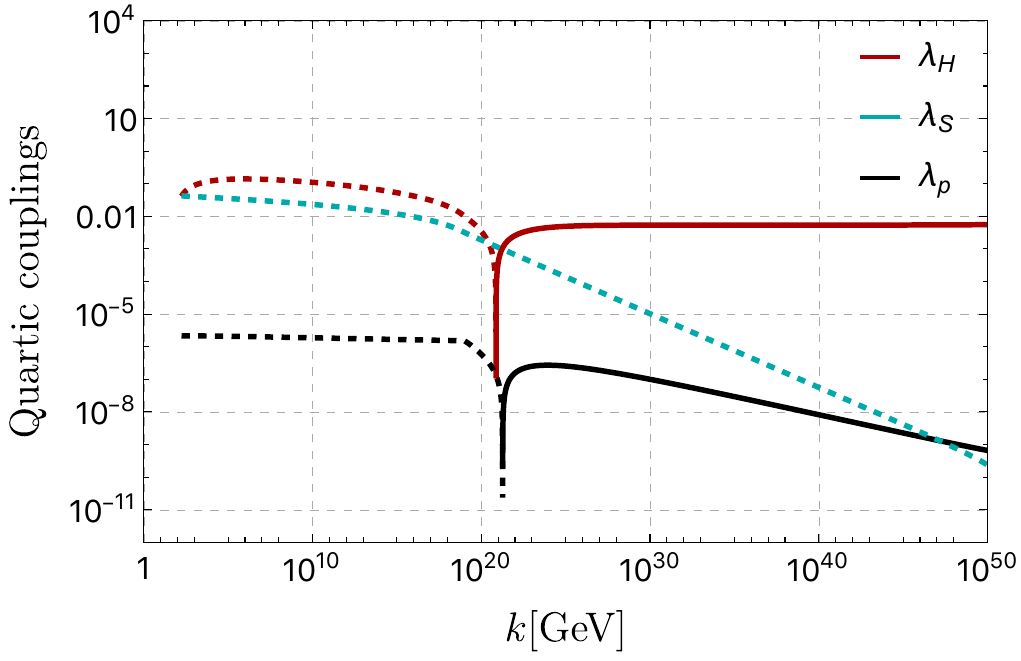} 
	\includegraphics[width=.325\linewidth] {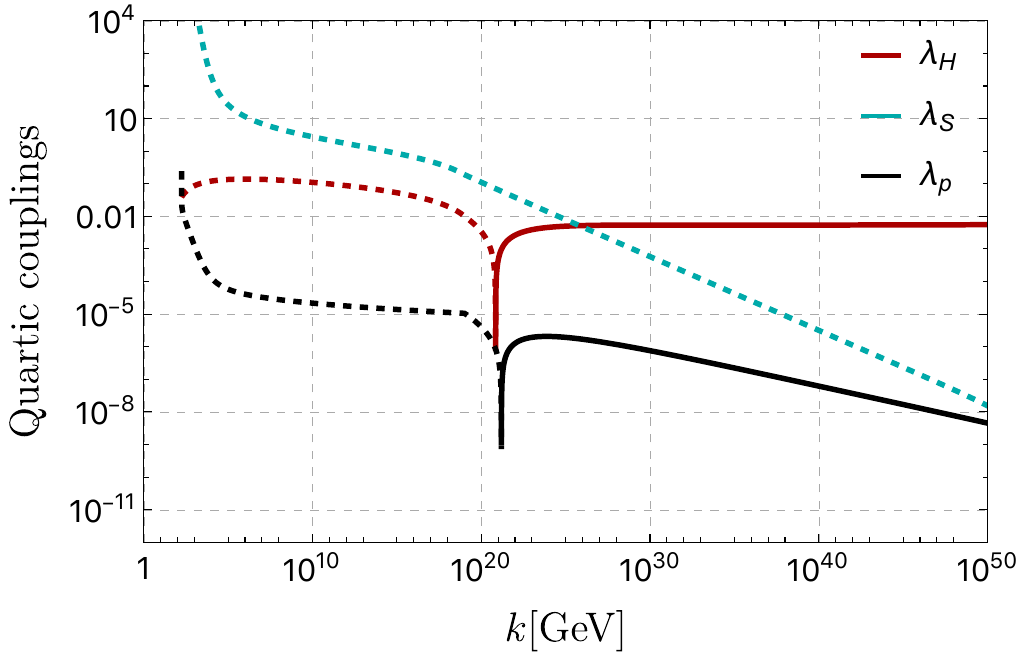} \\
	 \includegraphics[width=.325\linewidth] {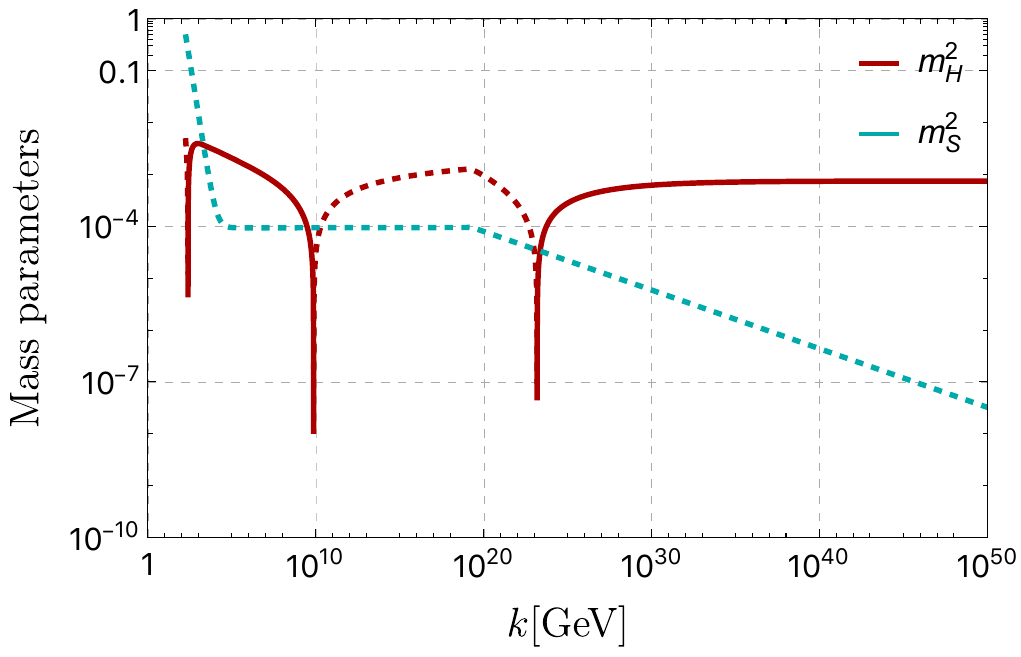}	
	\includegraphics[width=.325\linewidth] {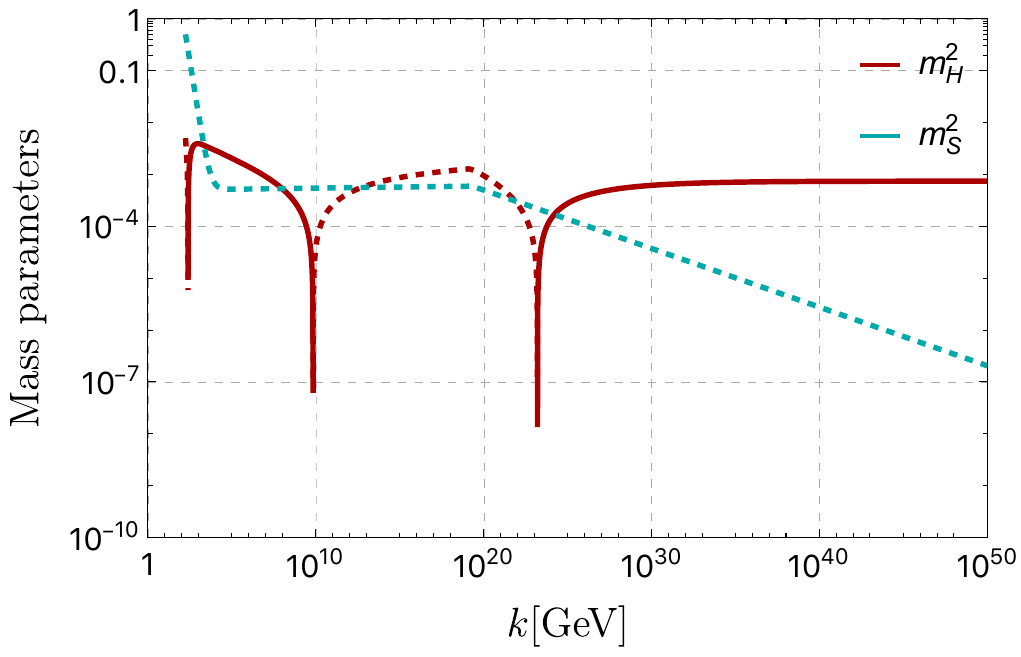}	
	\includegraphics[width=.325\linewidth] {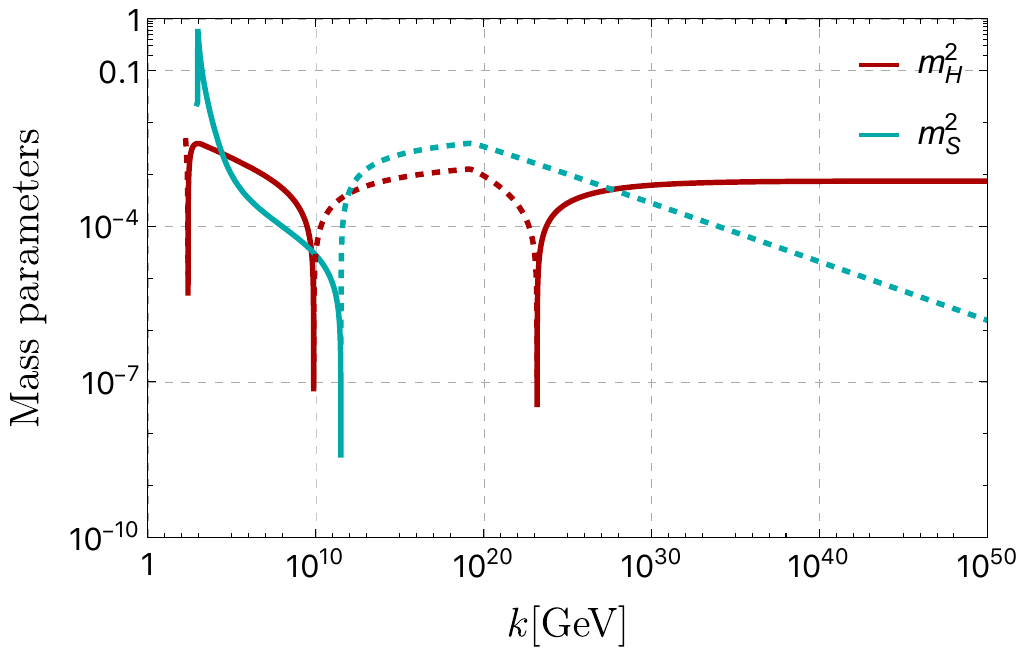}
	\caption{RG trajectories of various couplings in the $ \U(1)_\D $ DM model for different IR values of the dark gauge coupling, $g_{\D}(M_V=911~\textmd{GeV})=0.10$ (left), $0.25$ (middle) and $0.66$ (right panels). The top panels depict top-Yukawa and gauge couplings, where the RG flow of the bottom-Yukawa coupling is omitted and  we employ the normalization $g_1=\sqrt{5/3}g_\Y$. 
 The middle panels illustrate scalar couplings, with negative values (dashed lines) indicating unstable potential in both the Higgs and dark scalar directions, posing a challenge to asymptotic safety. Finally, the bottom panels display the dimensionless mass parameters of the Higgs and the dark scalar.
    }                  
	\label{fig:running-couplings}
\end{figure*}

 In Fig.~\ref{fig:running-couplings}, we plot the RG trajectories for various couplings in our model.
In the first row, we show the gauge and Yukawa couplings, which show a SM-like behavior for the SM gauge couplings. The dark gauge coupling is asymptotically free (which is a crucial difference to the setting without gravity) and decreases very slowly in the IR. The top Yukawa coupling increases towards the IR from its fixed-point value. Its predicted value in the IR is too low compared to measurements \cite{CMS:2018uxb,ATLAS:2018mme}. This is different from the result in \cite{Eichhorn:2017ylw}, where the predicted value was compatible with experiment. The difference between the two settings are the dark sector degrees of freedom. These impact the top Yukawa coupling indirectly: they change the gravitational fixed-point values, which in turn changes the top Yukawa fixed-point value. 

In the second row of Fig.~\ref{fig:running-couplings}, we show RG trajectories corresponding to the quartic couplings $\lambda_\text{H}$, $\lambda_\text{S}$ and $\lambda_\text{p}$. All quartic couplings become negative in the IR. This is caused by gauge field fluctuations, encoded in the  $g_\text{D}^4$-term in the beta function $\beta_{\lambda_\text{S}}$. Since the term proportional to $g_\text{D}^4$ has a positive coefficient, it drives the quartic coupling towards negative values in the IR.
In the beta functions for $\lambda_\text{H}$ and $\lambda_\text{p}$, there is a competition between positive and negative terms. In the UV, the negative terms dominate, such that $\lambda_\text{H}$ and $\lambda_\text{p}$ flow towards positive values.
Further in the IR, the positive terms in $\lambda_\text{H}$ and $\lambda_\text{p}$ start to dominate over the negative ones, thus pushing $\lambda_\text{H}$ and $\lambda_\text{p}$ towards negative values. The transition from positive to negative values of $\lambda_\text{H}$ and $\lambda_\text{p}$ happens around $k = 10^{21}\,\text{GeV}$.  All quartic couplings remain negative below this scale.

Furthermore, we have verified that the Higgs and dark quartic couplings flow towards negative IR values even when the mass parameters are excluded from the system of RG equations. The main difference is that, in this case, the portal coupling vanishes along the entire RG trajectory.

\begin{figure*}[t]
	\hspace*{-0.2cm}
	 \includegraphics[width=.33\linewidth] {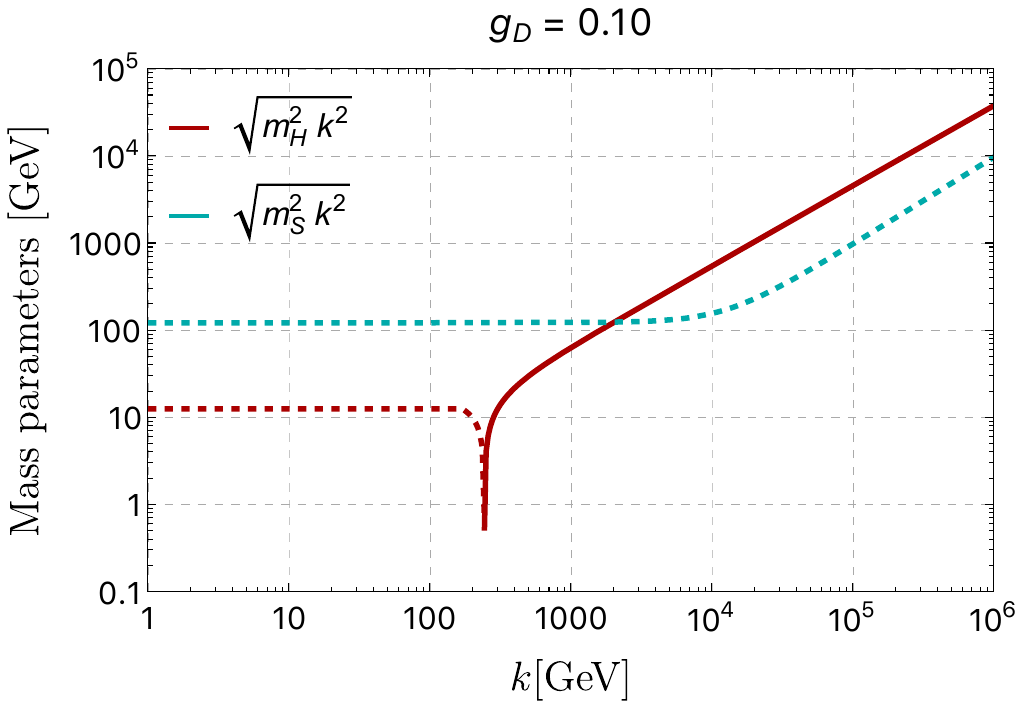}
	 \includegraphics[width=.33\linewidth] {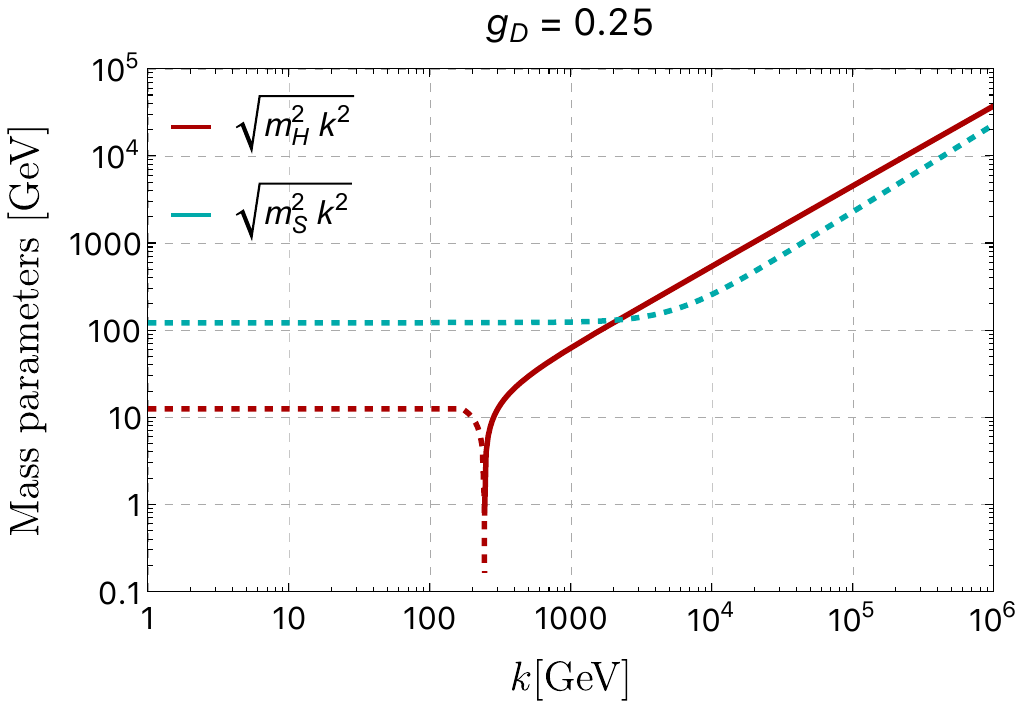}	
	\includegraphics[width=.33\linewidth] {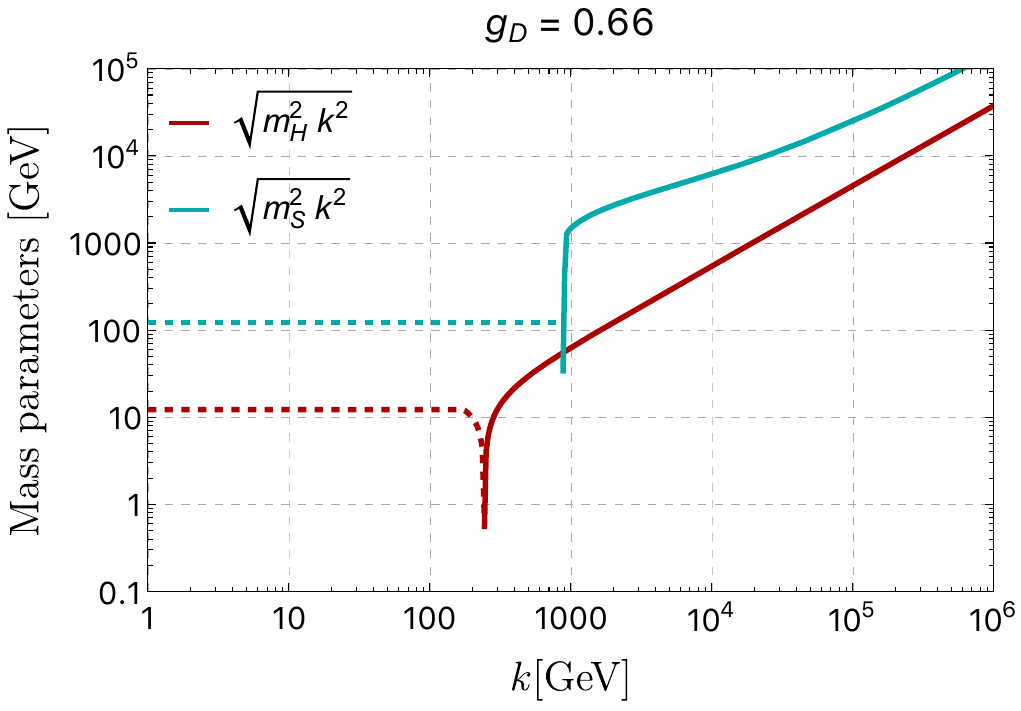}	
	\caption{RG trajectories of the dimensionful mass parameters ($ \sqrt{m_i^2 k^2} $ with $ i=\H,\S $) for various IR values of the dark gauge coupling, $g_{\D}(M_V=911~\textmd{GeV})=0.10$ (left), $0.25$ (middle) and $0.66$ (right panel).}                   
	\label{fig:mass parameters}
\end{figure*}

In the third row of Fig.~\ref{fig:running-couplings}, we plot the RG trajectories for the dimensionless mass parameters $m_\text{H}^2$ and $m_\text{S}^2$ (see also Fig.~\ref{fig:mass parameters} for RG trajectories of the dimensionful mass parameters ($ \sqrt{m_i^2 k^2} $ with $ i=\H,\S $) in the IR region).
For the Higgs mass parameter, we obtain the same behavior for all the benchmark values of the dark gauge coupling $g_\text{D}$. As we see, the running of $m_\text{H}^2$ oscillates between positive and negative values. We select its IR value according to Table.~\ref{table:infrared-values} as an attempt to obtain spontaneous symmetry-breaking in the IR. 
The dark mass parameter departs from its fixed point value $(m_\text{S}^2)_* = 0$ towards negative values. It remains negative all the way to the deep IR for the benchmark values $g_\text{D}(k = 911\,\text{GeV}) = 0.10$ and $g_\text{D}(k = 911\,\text{GeV}) = 0.25$. For $g_\text{D}(k = 911\,\text{GeV}) = 0.66$, $m_\S^2$ becomes positive in the IR.

The overall picture we obtain is that starting from an asymptotically safe trans-Planckian regime one can connect the gauge couplings to phenomenologically viable values in the IR. There are, however, two problems with other couplings. 
First, the top-quark Yukawa coupling is predicted  to be significantly smaller than the measured value. Because this prediction also serves as an upper bound for top-Yukawa values of fixed point (i), it also puts fixed point (i) in tension with experiment.
Second, the quartic scalar couplings run towards negative IR values, which results in a perturbatively unstable scalar potential.\footnote{In our setting, we cannot exlude that a nonperturbative IR setting may be phenomenologically viable and feature a stable potential in the presence of higher-order terms.}
This is a strong indication that the $\text{U}(1)_\text{D}$ DM model explored in \cite{Frandsen:2022klh} is incompatible with asymptotically safe quantum gravity. We emphasize that this conclusion is not just a consequence of the specific benchmark values used in our analysis. The instability of the scalar potential is a general consequence of the positive contribution involving gauge couplings to the $4^\text{th}$ power in the beta functions for the quartic couplings. Therefore, different benchmark values of $g_D(k = 911\,\text{GeV})$ can change quantitative details, but without changing our qualitative results.

\subsection{Constraints from asymptotic safety on $\SU(2)_\D$ dark matter}

The RG flow obtained with the $\text{SU}(2)_\text{D}$ DM model shares many of the features with the $\text{U}(1)_\text{D}$ model discussed in the previous section.
Nevertheless, there are important differences:
\begin{itemize}
    \item First, the non-Abelian dark gauge coupling $g_\text{D}$ is asymptotically free even in the absence of gravity. This reduces the number of fixed-point candidates, as $g_{{\rm D},\,\ast} = 0$ is the only possibility of UV completion for the non-Abelian dark gauge coupling. In this case, the possible fixed-point candidates share the same feature as the candidates (i)-(iv) in Table.~\ref{table:fixed-points}.
    \item Second, using the matter content of $\text{SU}(2)_\text{D}$ DM model, one obtains the fixed-point values $G_{\ast}=1.88$ and $\Lambda_{\ast}=-2.19$.
    At this fixed point, the value for the cosmological constant lies outside the viable region for UV completion of Yukawa couplings ($\Lambda \lesssim -3.3$ \cite{Eichhorn:2017ylw}), \ie, the Yukawa couplings vanish at the Planck scale. 
\end{itemize}

In principle, one can argue that the last point already indicates the incompatibility between the $\text{SU}(2)_\text{D}$ DM model with asymptotically safe quantum gravity. However, the fixed-point values of the gravitational couplings are subject to large systematic uncertainties, and it is interesting to explore the qualitative features of the RG flow independently of specific values of $G_{\ast}$ and $\Lambda_{\ast}$.

We thus explored RG trajectories obtained by integrating the flow with benchmark values for $G_{\ast}$ and $\Lambda_{\ast}$ lying inside the viable region for UV completion in the Yukawa sector. We focused on trajectories with boundary conditions corresponding to a UV fixed point with similar features as the fixed-point candidate (iii) in Table.~\ref{table:fixed-points}. The resulting trajectories are similar to the ones obtained in the previous section. 

In particular, the RG flow drives the quartic scalar couplings to negative values in the IR, resulting in a perturbatively unstable scalar potential. Again, this is a consequence of positive contributions to the beta functions of the quartic scalar couplings that are proportional to the gauge couplings. This result constitutes a strong indication of the incompatibility between the $\text{SU}(2)_\text{D}$ DM model and asymptotically safe quantum gravity.

\section{Conclusion}\label{sec:conclusions}

This paper demonstrates the predictive power of asymptotic safety in testing concrete DM models, in this case vector DM, which are phenomenologically not excluded. 

There are \textit{a priori} several ways in which asymptotic safety can be incompatible with a given phenomenological model.

First, there might not be any asympotically safe fixed point in the model together with quantum gravity. This is not the case here, where, in fact, there are several fixed points at which different subsets of the couplings are nonzero.

Second, a fixed point can have irrelevant directions. Each irrelevant direction generates a prediction for a coupling (or combination of couplings) in the infrared. These predictions may disagree with phenomenologically allowed or experimentally measured values. This is the case here, where the top-quark Yukawa coupling is predicted for some fixed points (and bounded from above by others) and comes out smaller than the value inferred from experiment. There is, however, a caveat to ruling out a model in this way and that caveat is due to systematic uncertainties. These arise from the use of a truncation (and thus the neglecting of higher-order interactions which are generically present at an asymptotically safe fixed point and impact the beta functions within the truncation), but also from the use of Euclidean signature in the RG calculations. Although attempts can be made  \cite{Eichhorn:2017ylw,Pastor-Gutierrez:2022nki}, these systematic uncertainties are difficult to estimate. If we are very conservative about the size of systematic uncertainties, ruling out the model robustly based on the too-low value of the top-quark mass may not be possible. If we are somewhat more optimistic about the size of systematic uncertainties, the present paper provides an explicit example of the idea put forward in \cite{Eichhorn:2017ylw}: dark degrees of freedom change gravitational fixed-point values and thus result in changes of all predicted SM couplings, even if the corresponding degree of freedom (here, the top quark) is not coupled to the dark sector. Furthermore, we emphasize that the DM models considered in this paper are only ruled out in the near-perturbative regime and it is in principle not excluded that a very different strongly coupled UV completion exists once we go beyond the approximations adopted.

Third, a model may rely on a dynamical mechanism, with spontaneous symmetry breaking as one important example. Whereas spontaneous symmetry breaking in phenomenological models is usually built in by fiat by making assumptions about the scalar potential, this is no longer an option in an asymptotically safe setting. There, the scalar potential at low energies arises as a combination of the UV initial conditions from asymptotic safety with the effect of quantum fluctuations at all scales down to the IR. The UV initial conditions are subject to the free-parameter count from asymptotic safety and typically only the mass parameters remain as free parameters. At all scales below, the contribution that arises from quantum fluctuations is completely fixed in terms of the other couplings in the model. In our case, these restrictions are sufficient to rule the model out, because most fixed points result in either perturbatively unstable or symmetry-broken potentials in the UV. At the remaining fixed points, gauge fluctuations in the dark sector drive the quartic coupling towards negative values. Because there are no free parameters in the scalar potential that could offset this effect, the resulting potential is not bounded from below within the perturbative regime.\\
We stress that this last criterion, in contrast to the second one, is a \emph{qualitative}, not a \emph{quantitative} one. Thus, it is less sensitive to the systematic uncertainties of our study, because it only relies on well-established, universal \emph{signs} of terms in beta functions.

Because of the qualitative and general nature of the criterion that rules out these particular DM models, we conjecture that vector DM models with only vectors and scalars in the dark sector are generally not viable in asymptotic safety in a near-perturbative regime. If fermions are added with a large enough Yukawa coupling, they may stabilize the scalar potential, see \cite{Reichert:2019car} in the context of DM and \cite{Eichhorn:2023gat} for work in the context of cosmic strings. We highlight that exploring vector dark models beyond the perturbative regime, where additional, canonically irrelevant, interactions may be relevant, is an interesting subject for further studies.

This result adds further to the evidence that models with a Higgs portal to the dark sector are either strongly constrained \cite{Eichhorn:2020kca,Eichhorn:2020sbo} or fully ruled out \cite{Eichhorn:2017als} in asymptotic safety, depending on the matter and interaction content of the dark sector. This motivates to look elsewhere for viable DM models. It is already known that axionlike particles are also constrained \cite{deBrito:2021akp}, putting another popular DM candidate under (theoretical) pressure. 

We thus propose that a way towards finding viable models of DM in asymptotic safety is to follow up on results which show which interaction structures are generically viable in asymptotic safety. A promising candidate could be strongly coupled fermionic sectors in which symmetry breaking triggers the formation of massive bound states. Such composite models with dark sectors may fit into the asymptotic safety paradigm \cite{Frandsen:2023vhu}, because it is known that i) gravity does not trigger bound-state formation, so that bound states less massive than the Planck scale may be viable \cite{Eichhorn:2011pc,Meibohm:2016mkp,Eichhorn:2017eht,deBrito:2020dta}, ii) gravity generates nonvanishing four-fermion interactions which may then be driven to criticality by a non-Abelian gauge interaction \cite{deBrito:2023} and iii) the fermion mass parameter generically remains a free parameter \cite{Eichhorn:2016vvy} so that even in nonchiral fermion systems bound-state formation may be achieved. If indeed composite models with dark sectors are viable in asymptotic safety, then a dark gauge interaction is present, but, unlike in the setting of this paper, does not supply the dark-matter candidate itself. 

\section*{Acknowledgments}

The authors thank Álvaro Pastor-Gutiérrez, Aaron Held and Manuel Reichert for fruitful discussions. G.P.B. and A.E. are supported by VILLUM FONDEN under Grant No. 29405. A.F.V. acknowledges funding by CNPq under the Grants No.~140968/2020-2 and 200442/2022-8. M.T.F. and M.R. acknowledge partial funding from The Independent Research Fund Denmark, Grants No. DFF 6108-00623 and No. DFF 1056-00027B, respectively. M.E.T. acknowledges funding from Augustinus Fonden, Application No. 22-19584, to cover part of the expenses associated with visiting the University of Helsinki for half a year. A.F.V. thanks  CP3-Origins at the University of Southern Denmark for the extended hospitality.

\appendix
\section{Gauge-fixed action for the SM-gravity subsystem}\label{App::gauge-fixed_action}

The gauge-fixed pure gravity sector of our truncated flowing action is given by the Einstein-Hilbert action
\al{
	\Gamma_{k,\textmd{grav}}&=\frac{1}{16 \pi \GN}\int_x\! \left(2\bar{\Lambda}-R(g)\right) \nn\\[1ex]
	&+\frac{1}{2\alpha}\int_{\bar{x}} \bar{g}^{\mu\nu} \mathcal{F}_{\mu}[h;\bar{g}]\mathcal{F}_{\nu}[h;\bar{g}]+ S_{\textmd{ghosts}}^{\rm grav},
}
where $R$ is the Ricci scalar and $\GN$ and $\bar{\Lambda}$ are the scale-dependent dimensionful Newton coupling and cosmological constant, respectively. Their dimensionless counterparts are obtained through $\GN(k)=k^{-2}G(k)$ and $\bar{\Lambda}(k)=k^2\Lambda(k)$.

Local coarse-graining techniques demand the introduction of a nondynamical background metric $\bar{g}_{\mu\nu}$. The (full) metric is expanded into an Euclidean background $\bar{g}_{\mu\nu}=\delta_{\mu\nu}$ and a dynamical (not necessarily small) fluctuation piece $h_{\mu\nu}$ as
\begin{equation}
	g_{\mu\nu}=\delta_{\mu\nu}+Z_h^{1/2}(32 \pi k^{-2} G(k))^{1/2}h_{\mu\nu}
\end{equation}
where $Z_h$ is the wave function renormalization factor for the graviton $h_{\mu\nu}$. The choice of a flat background is sufficient in order to compute the RG-flow of curvature-independent, matter couplings.

The linear gauge-fixing function is
\begin{equation}
	\mathcal{F}_{\mu}[h;\bar{g}]=\sqrt{2}Z_h^{1/2}\left(\delta^\alpha_\mu \bar{g}^{\nu\beta}-\frac{1+\beta}{4}\delta^\nu_\mu \bar{g}^{\alpha\beta}\right)\bar{\nabla}_\nu h_{\alpha\beta},
\end{equation}
where $\alpha$ and $\beta$ are gauge-fixing parameters. Here, $\bar{\nabla}_\mu$ stands for the spacetime covariant derivative defined with respect to the background metric. The Landau-gauge limit, $\alpha \rightarrow 0$, is adopted. The corresponding Faddeev-Popov ghost term $S_{\textmd{ghosts}}^{\rm grav}$ is computed from the gauge-fixing function $\mathcal{F}_{\mu}[h;\bar{g}]$.

The gauge-fixed SM truncation, without the Higgs potential, reads
\begin{align}
	\Gamma_{k,\text{SM}}^0&= \frac{1}{4}\int_{x}  W^{a}_{\mu\nu}  W^{a,\mu \nu}+ \,\frac{1}{4}\int_{x} B_{\mu\nu}  B^{\mu \nu} +\, S^{\text{EW}}_{\gf}+S^{\textrm{EW}}_{\textrm{ghosts}}\nonumber\\[1ex]
	&\quad\,+\frac{1}{4}\int_{x} G^{a}_{\mu\nu}    G^{a,\mu \nu} +S^{\SU(3)}_{\gf}  +S^{\SU(3)}_{\textrm{ghosts}}\nonumber\\[1ex]
	&\quad\,+ \sum_{j=1,2,3 }  \int_{x}  \, i \bar \psi^{ \L/ \R}_{i\,, j} \slashed D \, \psi^{\L/\R}_{i\,, j} +\Gamma_{k,\textrm{Yukawa}}\nn\\[1ex]
	&\quad\,+ \int_{x} (D_{\mu}  \Phi_i)^{\dagger }( D^{\mu }\Phi_i) \,.
	\label{eq:SMAction}
\end{align}
Here $B_\mu$ is the hypercharge gauge field, $W^a_\mu$ are the weak gauge fields and $A_\mu^a$ are the gluon gauge fields, with $B_{\mu\nu}=\pt_\mu B_\nu-\pt_\nu B_\mu$, $W^a_{\mu\nu}=\pt_\mu W^a_\nu-\pt_\nu W^a_\mu+g_2 \epsilon^{abc}W^b_\mu W^c_\nu$  and $G^a_{\mu\nu}=\pt_\mu A^a_{\nu}-\pt_\nu A^a_\mu +g_3f^{abc}A^b_\mu A^c_\nu$ being their respective field-strengths.
The fermionic fields $\psi_{i,j}$ represent general quark $q_{i,j}$ and lepton $l_{i,j}$ doublet fields, with $i$ being the isospin index and $j$ labelling the generation, ranging over the whole families of quarks and leptons of the SM. Compactly, we have $\psi_{i,j}=(q_{i,j},l_{i,j})$. The covariant derivative of the fermionic fields reads
\begin{align}
	D_\mu\psi_{i,j}^I&=\partial_\mu\psi_{i,j}^I +\omega_\mu\psi_{i,j}^I -i g_\Y Y B_\mu \psi_{i,j}^I\nn\\
	&-i g_2 W_{a,\mu} T^{a}_{ik}\psi_{k,j}^I +ig_3 A_{\mu,b}t^{b}_{IJ} \psi_{i,j}^J.
\end{align}
 The quarks live in the fundamental representation of the color group $\SU(3)_{\rm C}$ with color indices $I,J=1,\,2,\,3$. The matrices $T^a$ and $t^a$ are the generators of $\SU(2)_{\textmd{L}}$ and $\SU(3)_{\rm C}$, respectively. The hypercharge values $Y$ are assigned according to each quark and lepton generation and their respective chiralities. The coupling of the fermionic fields with gravity is via the $\textmd{so}(4)$-valued spin-connection $\omega_\mu$. Furthermore, the coupling of the Higgs doublet with the EW gauge group is done via the covariant derivative
\begin{gather}
	D_\mu\Phi_i=\partial_\mu\Phi_i -\frac{i}{2}g_\Y B_\mu \Phi_i -ig_2 W_{a,\mu} T^{a}_{ij}\Phi_j.
\end{gather}

The gauge-fixing and the associated Faddeev-Popov ghost action for the EW and QCD sectors are, respectively,
\begin{align}
	S^{\text{EW}}_{\gf}&+S^{\textrm{EW}}_{\textrm{ghosts}}=\frac{1}{2\xi_\W}\int_{\bar{x}}\left(\pt_\mu W^{a,\mu} \right)^2+\frac{1}{2\xi_\B}\int_{\bar{x}}\left(\pt_\alpha B^{\alpha} \right)^2\nn\\[1ex]
	&+\int_{\bar{x}}\left(\bar{c}^{(2)}_a\pt_\mu\pt^\mu c^{(2)}_a-g_2\epsilon^{abc}\bar{c}^{(2)}_a \pt_\alpha\left(W_{c}{}^{\alpha}c^{(2)}_b\right)\right),
\end{align}
\begin{align}
	S^{\SU(3)}_{\gf}  &+S^{\SU(3)}_{\textrm{ghosts}}=\frac{1}{2\xi_\A}\int_{\bar{x}}\!\left(\pt_\mu A^{a,\mu} \right)^2\nn\\[1ex]
	&\int_{\bar{x}}\left(\bar{c}^{(3)}_a\pt_\mu\pt^\mu c^{(3)}_a-g_3f^{abc}\bar{c}^{(3)}_a \pt_\alpha\left(A_{c}{}^{\alpha}c^{(3)}_b\right)\right),
\end{align}
For the Yukawa sector, we consider an approximation where only the top and bottom quarks have nonvanishing (real) Yukawa couplings. The explicit action reads
\begin{align}
	\Gamma_{k,\rm Yukawa}&=\int_x y_t\left(\bar{q}^\L_{i,3} \Phi^i q^\R_{1,3}+\bar{q}^\R_{1,3}\Phi^{\dagger i} q^{\L}_{i,3} \right)\nn\\[1ex]
	&+\int_x y_b\left(\bar{q}^\L_{i,3} \tilde{\Phi}^i q^\R_{2,3}+\bar{q}^\R_{2,3}\tilde{\Phi}^{\dagger i} q^{\L}_{i,3} \right),
\end{align}
where the spinorial and color indices are omitted and $\tilde{\Phi}^i=i\sigma_2^{ij}\Phi^\dagger_j$ with $\sigma_2$ being the second Pauli matrix. The third generation of the right chirality $\SU(2)_\L$ quark doublet is explicitly given by $q^\R_{i,3}=(t^\R,\,b^\R)^{\rm T}$. Moreover, all the SM and beyond SM fields, including ghosts, are augmented by wave function renormalization factors, \ie, $\phi_{\rm SM}\mapsto Z_{\rm SM}^{1/2}\,\phi_{\rm SM}$ and $\phi_{\rm DM}\mapsto Z_{\rm DM}^{1/2}\,\phi_{\rm DM}$.


\bibliography{refs}

\end{document}